\documentclass[twocolumn,showpacs,preprintnumbers,amsmath,amssymb,
superscriptaddress, array]{revtex4}

\usepackage{bm}

\usepackage[dvips]{graphicx}

\advance\textheight by 0.23cm

\makeatletter

\input epsf
\usepackage{amsmath,amssymb,epsfig,placeins,subfigure,wrapfig,indentfirst}

\renewcommand{\thesubfigure}{ \thefigure (\alph{subfigure})}
\renewcommand{\p@subfigure}{}
\renewcommand{\@thesubfigure}{\thesubfigure:\hskip\subfiglabelskip}

\begin{document}
\title{Physics of the interior of a black hole with an exotic scalar matter}n

\author{Andrey Doroshkevich}
\affiliation{Astro Space Center, Lebedev Physical Institute,
Russian Academy of Sciences, Moscow, Russia}
\author{Jakob Hansen}
\affiliation{Korea Institute of Science and Technology Information, 335 Gwahak-ro, Yuseong-gu, Daejeon, 305-806, Korea}
\affiliation{Department of Physics, Waseda University, Okubo 3-4-1, Shinjuku,
Tokyo, Japan}
\author{Dmitriy Novikov}
\affiliation{Astro Space Center, Lebedev Physical Institute,
Russian Academy of Sciences, Moscow, Russia}
\affiliation{Astrophysics Group, Imperial College, Blackett
Laboratory, Prince Consort Road, London, SW7 2AZ, United Kingdom}
\author{Igor Novikov}
\affiliation{Astro Space Center, Lebedev Physical Institute,
Russian Academy of Sciences, Moscow, Russia} \affiliation{The
Niels Bohr International Academy, The Niels Bohr Institute,
Blegdamsvej 17, DK-2100 Copenhagen, Denmark}
\author{Dong-Ho Park}
\affiliation{Korea Institute of Science and Technology Information, 335 Gwahak-ro, Yuseong-gu, Daejeon, 305-806, Korea}
\affiliation{School of Physics \& Astronomy, Seoul National University, Seoul 141-747, Korea}
\author{Alexander Shatskiy}
\affiliation{Astro Space Center, Lebedev Physical Institute,
Russian Academy of Sciences, Moscow, Russia}

\date{\today}

\begin{abstract}
{\bf ABSTRACT} 
We use a numerical code to consider the nonlinear processes arising when a
Reissner-Nordstr\"om black hole is irradiated by an exotic scalar field (modelled as a free massless scalar field with an opposite sign for its energy-momentum tensor).
These processes are quite different from the processes arising in the case of the same black
hole being irradiated by a pulse of a normal scalar field.
In our case, we did not observe the creation of a spacelike strong singularity in the T-region of the space-time.
We investigate the antifocusing effects in the gravity field of the exotic scalar
field with the negative energy density and the evolution of the
mass function.
We demonstrate the process of vanishing of the black hole when it is irradiated by a strong pulse of an exotic scalar field.
\end{abstract}


\maketitle

\section{Introduction}
\label{Introduction}

The internal structure and physics of black holes (BH) has been the
subject of researches during many years \cite{Penrose1, Dorr1,
Namara1, Namara2, Gursel1, Gursel2, Matzner1, Chandra1,
Goldwirth87, Poisson90, Ori91, Ori92,  Gnedin93, Bonanno94,
Brady95, Droz96, Burko97b, Burko97c, Hod97,  Burko98c, Hod98,
Hod98b, Burko99-1, Burko99, Ori99, Ori99b, Burko02, Berger02,
Burko02b, Oren03, Hamilton04a, Hamilton04b, Hansen1,
Dorr2,Krasnikov1,dop-9,Dokuchaev04}. A powerful tool for these
investigations is to consider a spherical, charged, nonrotating BH which
is nonlinearly perturbed by a selfgravitating scalar field.
This toy BH model is not very realistic but it shares many properties, including causal
structure, with the more realistic rotating BHs.

This toy model has been used in the paper \cite{Hansen1} to
analyze the physics of the interior of a BH in the case of
irradiation by a normal massless scalar field.

The observational apparent acceleration of the universe suggests the
presence of a matter field which violates at least the strong and perhaps
also the weak energy condition and one of the simplest examples of such
energy-conditions-violating matter is a free scalar field with a reversed
sign of its energy-momentum tensor, we call this an exotic scalar field. Such an exotic scalar field has previously been used in 
many works for construction of wormholes in cosmological models with nontrivial topologies, 
see for example~\cite{Bronnikov73, Bronnikov10, Dorr2}. It would thus be interesting to explore
possible implications of such a field to black hole physics.

The goal of this paper is to perform such an analysis, where we
will model the exotic matter as described above.
We will see that the physics of the interior of a BH which is nonlinearly perturbed by such an exotic scalar
field is quite different from the physics of a BH perturbed by normal scalar field.

Rather recently, some aspects of the problem of the internal
structure of a BH with unusual scalar fields have been discussed
in the papers~\cite{Hong2008, Dong-han2008, Gonzalez2009}, see
also references therein.

In the paper~\cite{Hong2008} the authors discussed the interior
structure of a BH arising as a result of the collapse of a
spherical charged scalar field shell, taking into account the
Hawking radiation. The problem was treated numerically.

In the paper~\cite{Dong-han2008} some problems of the inner
structure of a BH were discussed using an analytical approach.

Finally, in the paper~\cite{Gonzalez2009}, the authors explored
numerically the accretion of a phantom scalar field with a
nonzero potential onto a noncharged BH. Especially the variation
of the BH area were investigated.

Our approach is different from the above papers. We will address the following points:

$\bullet$ What physical processes arise in the case of irradiation
of a charged BH by an exotic scalar field and what is the difference from the
case of irradiation by a normal scalar field.

$\bullet$ What kind of singularities arise inside of a charged BH in the
case of irradiation by an exotic scalar field.

$\bullet$ How is the the charged BH transformed into an object without horizons in the case of the BH
being irradiated by stronger pulses of exotic scalar radiation?

$\bullet$ What are the proper Penrose diagrams for both
surviving and nonsurviving (charged) BHs in the case of irradiation by
an exotic scalar field.

We will study the nonlinear processes inside the BH using a
stable, second order accurate numerical code with adaptive mesh
refinement capabilities. This code was described and tested in
details in \cite{Dorr2}, see also the Appendix~\ref{sec:app1}.

This paper is organized as follows:

In section~\ref{sec2}, Einstein's equations are written for a
spherical BH with a fixed magnetic charge and massless scalar
fields, both normal and exotic. In section~\ref{sec3}, we formulate
the initial values for our computations. In section~\ref{sec4}, we
discuss the mass function and some nonlinear processes inside a
BH. In section~\ref{sec5}, we discuss the problem of the origin of
strong singularities in a BH. In section~\ref{sec6}, we describe our
numerical model. In section~\ref{sec7}, we remind of the physical
processes inside a charged BH irradiated by a pulse of the normal
scalar field. In section~\ref{sec8}, we analyze the results of our
computations which describe the physics of the interior of a BH with a
magnetic field irradiated by a pulse of exotic scalar radiation.
In section~\ref{sec9}, we discuss the case of irradiation of the BH
by both normal and exotic radiation pulses and we summarize our conclusions in
section~\ref{concl}. Finally, in appendix A, we presents results of
convergence tests of the numerical code.

\section{Field equations for the spherical model}
\label{sec2}

We wish to investigate the evolution of a spherical BH with a
fixed magnetic or electric charge $q$ (i.e. Reissner-Nordstr\"om
metric) under the action of pulses of selfgravitating,
massless scalar fields, both normal $\Phi$ (with a positive energy
density, $\varepsilon >0$) and exotic $\Psi$ (with a negative
energy density, $\varepsilon <0$). The general equations for the
analysis were written in \cite{Dorr2}.

The line element in double-null coordinates can be written as
\begin{equation}
\label{ds2}
ds^2 = -2 e^{2 \sigma (u,v)} du\, dv + r^2 (u,v)
d\Omega^2,
\end{equation}
where $\sigma (u,v)$ and $r(u,v)$ are functions of the null
coordinates $u$ and $v$ (in- and out-going respectively). The
nonzero components of the Einstein-tensor are:
\begin{subequations}
\label{eq:4}
\begin{eqnarray}
 G_{uu} &=& \frac{4 r_{,u} \sigma_{,u} -2r_{,uu} } {r}\label{eq:Guu}\\
 G_{vv} &=& \frac{4 r_{,v} \sigma_{,v} -2r_{,vv} } {r}\label{eq:Gvv}\\
 G_{uv} &=& \frac{e^{2\sigma} + 2r_{,u} r_{,v} + 2 r r_{,uv}}{r^2}\\
 G_{\theta\theta} &=& -2 e^{-2\sigma} r (r_{,uv} + r \sigma_{,uv})\\
 G_{\varphi\varphi} &=& -2 e^{-2\sigma} r \sin^2\theta (r_{,uv} + r
\sigma_{,uv})
\end{eqnarray}
\end{subequations}
The energy-momentum tensor can be written as a sum of
contributions from the exotic scalar field $\Psi$ (with the
negative energy density), from the normal scalar field $\Phi$ (with the
positive energy density) and from the ordinary electro-magnetic field, i.e. :
$T_{\mu\nu}= T_{\mu\nu}^{\Psi} +
T_{\mu\nu}^{\Phi}+T_{\mu\nu}^{em}$ :
\begin{equation}
\label{eq:5}
   T_{\mu\nu}^{\Psi} =\frac{-1}{4\pi} \begin{pmatrix}
   \Psi^2_{,u} & 0 & 0 & 0 \\
   0 & \Psi^2_{,v} & 0 & 0 \\
   0 & 0 & r^2 e^{-2\sigma}\Psi_{,u}\Psi_{,v} & 0 \\
   0 & 0 & 0 & r^2 \sin^2\theta\, e^{-2\sigma}\Psi_{,u}\Psi_{,v}
   \end{pmatrix}
   \end{equation}

\begin{equation}
\label{eq:6}
   T_{\mu\nu}^{\Phi} =\frac{+1}{4\pi} \begin{pmatrix}
   \Phi^2_{,u} & 0 & 0 & 0 \\
   0 & \Phi^2_{,v} & 0 & 0 \\
   0 & 0 & r^2 e^{-2\sigma}\Phi_{,u}\Phi_{,v} & 0 \\
   0 & 0 & 0 & r^2 \sin^2\theta\, e^{-2\sigma}\Phi_{,u}\Phi_{,v}
   \end{pmatrix}
   \end{equation}

\begin{equation}
\label{em-eq}
   T_{\mu\nu}^{em} =\frac{q^2}{8\pi r^4} \begin{pmatrix}
   0 & e^{2\sigma} & 0 & 0 \\
   e^{2\sigma} & 0 & 0 & 0 \\
   0 & 0 & r^2 & 0 \\
   0 & 0 & 0 & r^2 \sin^2\theta
   \end{pmatrix}
   \end{equation}
The $u-u$. $v-v$, $u-v$ and $\theta-\theta$ components of the
Einstein equations (with $c=1$, $G=1$) respectively are:
\begin{eqnarray}
  r_{,uu} - 2\, r_{,u}\,\sigma_{,u} - r\, \left(\Psi_{,u}
\right)^2 + r\,
\left(\Phi_{,u} \right)^2 &=& 0  \label{eq:7}\\
  r_{,vv} - 2\, r_{,v}\,\sigma_{,v} - r\, \left(\Psi_{,v}
\right)^2 + r\,
\left(\Phi_{,v} \right)^2 &=&0   \label{eq:8}\\
  r_{,uv} +\frac{r_{,u}
r_{,v}}{r}+\frac{e^{2\sigma}}{2r}\cdot\left(1-\frac{q^2}{r^2}\right) &=& 0
\label{eq:9}\\
 \sigma_{,uv} - \frac{r_{,v} r_{,u}}{r^2} -
\frac{e^{2\sigma}}{2r^2}\cdot\left(1-\frac{2q^2}{r^2}\right) & & \nonumber\\
 -\Psi_{,u}\Psi_{,v}  + \Phi_{,u}\Phi_{,v}  &=& 0   \label{eq:10}
\end{eqnarray}
The scalar fields satisfy the Gordon-Klein equation
$\nabla^{\mu}\nabla_{\mu}\Psi = 0$ and
$\nabla^{\mu}\nabla_{\mu}\Phi = 0$, which in the metric
\eqref{ds2} become:
\begin{eqnarray}
& & \Psi_{,uv} + \frac{1}{r} \left( r_{,v}\Psi_{,u} +
r_{,u}\Psi_{,v} \right) =
0 \label{eq:11a}\\
& & \Phi_{,uv} + \frac{1}{r} \left( r_{,v}\Phi_{,u} +
r_{,u}\Phi_{,v} \right) = 0 \label{eq:11b}
\end{eqnarray}
Equations \eqref{eq:9}-\eqref{eq:11b} are evolution equations
which are supplemented by the two constraint equations
\eqref{eq:7} and \eqref{eq:8}. It is noted that none of these
equations depend directly on the scalar fields $\Psi$ and $\Phi$
but only on their derivatives, i.e. the derivative of the scalar
field is a physical quantity, while the absolute value of the
scalar field itself is not. Specifically we note the
$T_{uu}^{\Psi} = - (\Psi_{,u})^2 / (4\pi)$  and $T_{vv}^{\Psi} = -
(\Psi_{,v})^2 / (4\pi)$ components of the energy-momentum tensor
\eqref{eq:5} and $T_{uu}^{\Phi} = (\Phi_{,u})^2 / (4\pi)$  and
$T_{vv}^{\Phi} = (\Phi_{,v})^2 / (4\pi)$ components of the
energy-momentum tensor \eqref{eq:6}, which are part of the
constraint equations. Physically $T_{uu}$ and $T_{vv}$ represents
the flux of the scalar field through a surface of constant $v$ and
$u$ respectively.

\section{Initial-value problem}
\label{sec3}

We wish to evolve the unknown functions $r(u,v)$,
$\sigma (u,v)$, $\Phi (u,v)$ and $\Psi (u,v)$ throughout some
computational domain. We do this by following the approach of
\cite{Burko97c, Burko02b, Burko97, Burko99-1, Burko98c, Burko99,
Burko02, Hansen1, Pretorius04, Dorr2} to numerically integrate the four
evolution equations \eqref{eq:9}-\eqref{eq:11b} along the characteristics.
These equations form a well-posed initial-value problem and in
the double-null coordinate we use,
it is most natural to use the characteristic initial-value formulation in which the
initial values of the unknown functions are specified on two initial null segments,
namely an ingoing (${v=v_0 = constant}$) and an outgoing (${u=u_0
= constant}$) segment \cite{Burko97}.

Our code is a free evolution code, i.e. we impose the constraint equations
\eqref{eq:7} and \eqref{eq:8} only on the initial segments, consistency
of the evolving fields with the constraint equations is then
ensured via the contracted Bianchi identities \cite{Burko97}. However
we use the constraint equations throughout the domain of
integration to check the accuracy of the numerical simulation.

We are free to choose the distribution
of the $\Phi$ and $\Psi$-fields on the initial null segments, our
choices are described in section \ref{sec6}. We are
also free to choose $r(u,v)$ on the initial surfaces, this merely
expresses the gauge freedom associated with the transformation
$u\rightarrow \tilde{u}(u), v\rightarrow \tilde{v}(v)$ (the line
element \eqref{ds2} and the equations \eqref{eq:7}-\eqref{eq:11b}
are invariant to such a transformation). Hence, the only variable
left for us to determine on the initial surfaces is $\sigma$. This
can easily be found by integrating the constraint equations eq.
\eqref{eq:7} and \eqref{eq:8}, which ensures that the constraint
equations are satisfied on the initial hypersurfaces. Specifically
on the outgoing $u=u_0$  hypersurface it is found by the integral:
\begin{equation}
\label{eq:sigma1} \sigma(u_0,v) = \sigma(u_0,v_0)
 + \int\limits_{v0}^{v}
\frac{r_{,vv} - r\, \left(\Psi_{,v} \right)^2 + r\,
\left(\Phi_{,v} \right)^2 }{ 2\, r_{,v}}dv
\end{equation}
where $\sigma(u_0, v_0)$ is an integration constant. We follow \cite{Burko02, Burko02b, Burko97, Hansen1, Dorr2} and set
$\sigma(u_0, v_0) = \ln\left(\frac{1}{\sqrt{2}}\right)$ throughout the paper.
The initial data along the ingoing surface, is set in a similar manner.

Hence, by specifying a distribution of the scalar
fields $\Phi$ and $\Psi$ on the initial null segments, choosing a
gauge, integration constant $\sigma(u_0, v_0)$, charge and mass parameters "$q$" and "$m$" (these relate to the gauge,
see eq. \eqref{mass1} and section \ref{sec6})\, we can specify
complete initial conditions on the initial null segments. Using
the numerical code from \cite{Dorr2}, we can then use the
evolution equations, eqs. \eqref{eq:9}-\eqref{eq:11b}, to evolve
the unknown functions along the characteristics, throughout the computational domain.
Our specific choices for initial conditions are described in sec. \ref{sec6}.

Finally, we wish to emphatize that a consequence of evolving along the characteristics in our
 double-null coordinate, spherical model, is that the specification of initial data along in- and outgoing null surfaces fully
  defines the initial-value problem. Thus, there is no need to specify any additional boundary conditions, the initial values
along the null surfaces \textit{are}
the boundary conditions. For further details on the numerical code, see \cite{Dorr2} and references therein.

\section{Mass function}
\label{sec4}

The evolution of the interior of a BH with a scalar field is
highly nonlinear. An important characteristic of this evolution
is the mass function. This function represents the total mass
(without the magnetic field) in a sphere of radius $r(u,v)$ (see
\cite{Poisson90,Burko97,Burko98c,Burko99}). In the
metric~(\ref{ds2}), the mass function is (see
\cite{Oren03,Burko02}):
\begin{equation}
\label{mass1}
m=\frac{r}{2}\left(1+\frac{q^2}{r^2}+\frac{2r_{,u}r_{,v}}{e^{2\sigma}}
\right)
\end{equation}
In the metric
\begin{equation}
\label{ds2-2} ds^2=g_{tt}\, dt^2+g_{rr}\, dr^2+r^2\, d\Omega^2
\end{equation}
the mass function $m$ has the form
\begin{equation}
\label{mass2} m=\frac{r}{2}\left(1+\frac{q^2}{r^2}-g_{rr}^{-1}
\right)
\end{equation}
There are two physical processes which can lead to a nonlinear change of
the mass function:

1. The mass $m$ inside a sphere can change because of the work of
the pressure forces acting on the surface of the sphere.

2. Mass inflation can occur~\cite{Poisson90}.

Both processes have been described in \cite{Hansen1}. We want to
emphasize that in the case of the exotic scalar field we should
take into account the negative energy density and negative radial
pressure of the exotic scalar field. Of course, in addition to the
above mentioned processes, there is also the trivial process of
matter flowing into the sphere. We will discuss the properties of
the mass function in our model in the subsequent sections.

One more important nonlinear process is the focusing effect caused by
the gravity of pulses of opposite fluxes of radiation (see
\cite{Hansen1}). In our case of radiation pulses with negative
energy densities, it is rather a defocussing effect. Its
manifestation will be discussed below.

\section{Strong singularity}
\label{sec5}
In the case of nonlinear perturbations of the initial BH by the
normal $\Phi$ scalar field, a strong spacelike singularity arises
in a T-region as a result of the nonlinear processes (see for
example \cite{Gnedin92, Gnedin93, Hansen1}).

We want to remind of the classification of $R$- and $T$-regions in
space-times with spherical symmetry \cite{Nov01, Novikov_old,
Frolov98}. The $R$-regions are regions where world-lines $r=const$
are timelike. The $T_{+}$-regions are regions where world-lines of
$r=const$ are spacelike and the direction to larger $r$ is in the
direction to the future. The $T_{-}$-regions are regions where
world-lines of $r=const$ are spacelike and the direction to smaller
$r$ is in the direction to the future. The $R$- and $T$-regions
are separated by apparent horizons which are defined by locations
where $\frac{dr}{dv} = 0$ or $\frac{dr}{du} = 0$.

The strong singularity under discussion exists together with a
weak null singularity (instead of the inner horizon of a
Reissner-Nordstr\"om BH~\cite{Ori92}).

In the case of nonlinear perturbations of the initial BH by the
exotic $\Psi$-field the situation is quite different.

First of all, the analysis in \cite{Novikova2009} have demonstrated
that in the case of a $\Psi$ field in the vicinity of the
singularity $r=0$, we can neglect the influence of the exotic scalar
field $\Psi$ comparative to the influence of the magnetic field.
This means that the singularity $r=0$ is of a Reissner-Nordstr\"om
type. It is a timelike singularity and it is in an R-region. A
spacelike strong singularity $r=0$ in a $T_{-}$-region is not
possible.

Secondly, we will see in the subsequent sections that under the action of
the irradiation by the exotic scalar field, the $T_{-}$-region
becomes smaller and smaller when the intensity of the irradiation
increases, even in the case when the BH still survives. We can see
practically all of the $T_{-}$-region in our computational domain and
definitely there are not any strong singularities in the $T_{-}$-region.
On the other hand, a strong timelike singularity of the
Reissner-Nordstr\"om type must exist in the R-region outside of our
computational domain.

We will consider the problems of the singularities in some detail in the following sections.

Finally, we wish to note that no special treatment is needed in order for our numerical code to successfully
handle the spacelike $r=0$ singularities that occur in our computations. The characteristic evolution of our numerical code means
that only points which are causally connected to a singularity can ``see'' the singularity, but since no points are causally connected
to a spacelike $r=0$ singularity, they pose no problem.
Timelike and coordinate ($r=0$) singularities, on the other hand, \textit{do} pose a problem in principle as
these points are causally connected to the rest of the space-time. In practice, however,
it is not a problem since we choose our initial conditions in such a way that they do not include the coordinate singularity
(cf. sec. \ref{sec6}) and any timelike singularity are outside of our computational domain.

For further details on the workings of our numerical code, please see \cite{Dorr2} and references therein.

\section{The model}
\label{sec6}
In the subsequent sections we investigate the fully nonlinear
processes in the spherical charged BH irradiated by a pulse of the
exotic scalar field $\Psi$ (and in section \ref{sec9} irradiated by both $\Psi$ and $\Phi$ pulses).
For that we use the results of our numerical simulations.
Our numerical code is thoroughly described
and tested in \cite{Dorr2}, here we just mention that the code uses Adaptive Mesh Refinement (AMR) in both $u-$ and $v-$coordinates and
is second order accurate. About the convergence of the code in
the case of the present model, see Appendix~\ref{sec:app1}.

Our choice of the initial values corresponds to the following
physical situation; There is a BH with a magnetic field and at
some distance from the horizon, at the initial moment, there is a
rather narrow spherical layer of an in-falling scalar field.

Throughout this paper, prior to the irradiation by the scalar
pulses, the BH has initial mass $m_0=1$ and charge $q=0.95m_0$.
For the simulations, the domain of integration as a rule is $5\le
v \le 30$, $0\le u\le 30$ and we choose a standard gauge in which
$r$ is linear in $v$ and $u$ on the initial null segments.
Specifically, our gauge choice is:
\begin{equation}
  \label{eq:2_10}
  r(u_0,v)=v \ \ \ \ \ \ \  r(u,v_0) = r(u_0,v_0) + r_{,u}\, u
\end{equation}
where $r_{,u}$ is determined from the mass function, eq. \eqref{mass1}, at the initial moment $u_0, v_0$, to ensure consistency.
We note that this computational domain does not include the $r=0$ coordinate singularity.

For our numerical experiments we specify various initial conditions.
Along the ingoing border $v_0=5$, the initial
values for all simulations correspond to the (Reissner-Nordstr\"om) BH solution with $m_0=1$ and $q=0.95m_0$
without any additional fields. Along the outgoing border $u_0=0$, we vary
the initial conditions in such a way as to imitate some physical
fluxes of the fields $\Phi$ and $\Psi$ into the charged BH.

We specify the pulses of the scalar fields in the form:
\begin{equation}
\label{Phi_v} \Phi_{,v} (u_0 ,v) = A_\Phi \sin^2 \left( \pi
\frac{v-v_{\Phi 0}}{v_{\Phi 1} - v_{\Phi 0}}\right),
\end{equation}
\begin{equation}
\label{Psi_v} \Psi_{,v} (u_0 ,v) = A_\Psi \sin^2 \left( \pi
\frac{v-v_{\Psi 0}}{v_{\Psi 1} - v_{\Psi 0}}\right),
\end{equation}
where $A_\Phi$ and $A_\Psi$, $v_{\Phi 0}$, $v_{\Phi 1}$, $v_{\Psi
0}$, $v_{\Psi 1}$ are
constants.\\
Outside the interval $(v_{\Phi 0},v_{\Phi 1})$ we have
$\Phi_{,v}(u_0,v)=0$.\\
Outside the interval $(v_{\Psi 0},v_{\Psi 1})$ we have
$\Psi_{,v}(u_0,v)=0$.

If the fluxes through the two initial surfaces $u_0$ and $v_0$ have
been chosen, as was described in section \ref{sec3}, all other initial
conditions are determined by our choice of gauge and the
constraint equations.

\section{Physical processes inside of the charged black hole irradiated by a
pulse of normal radiation}
\label{sec7}
We start from reminding of what happens if we irradiate a charged
BH with a pulse of the normal scalar field with a positive
energy density \cite{Burko97c, Burko02, Burko02b, Hod97, Hansen1}.

We described the corresponding processes in detail in~\cite{Hansen1}, here we summarize main effects:

1) When the positive energy pulse crosses the outer apparent
horizon (OAH), the horizon becomes bigger. When the pulse crosses
the inner apparent horizon (IAH1), it becomes smaller.

After the passage of the pulse, the mass function increases by an
amount corresponding to the energy contained within the pulse. These
effects are the trivial consequences of mass being pumped into the
BH by the initial pulse.

2) Next is the focusing effect. Any test photons propagating along
the null-geodesic lines $u=const$ and $v=const$ are under action of the gravity of the
fluxes $T_{vv}$ and $T_{uu}$ respectively. The last arises as a result of
the scattering of the ingoing $T_{vv}$ radiation by the space-time
curvature. In the subsequent evolution these $T_{vv}$ and $T_{uu}$ fluxes
are converted into one another due to the curvature of the space-time.
The gravity of $T_{vv}$ and $T_{uu}$ leads to a focusing effect on
the test photons. In the absence of the scalar radiation, outgoing
photons along $u=const$ slightly inside IAH1 will go to greater
$r$ as $v$ increases in the Reissner-Nordstr\"om solution. With the
existence of the scalar radiation, a similar outgoing ray will
now, because of the focusing effect, go to smaller $r$ and
generate a maximum ${\left.\frac{du}{dv}\right|_{r=const}=0}$ in
the ${u-v}$ coordinates. This corresponds to the position of the apparent
horizon. As a result the position of IAH1 changes dramatically. It
goes up sharply with increasing $v$ coordinate.

3) Now we consider the properties of the mass function. For late
(big) $u$ the mass function along the lines $u=const$ demonstrates
a sharp increase with growing $v$. This is the result of the
compression which tends to infinity. As a consequence, a strong
singularity $r=0$ arises and these lines, $u=const$, come to the
singularity $r=0$. The Kretschmann scalar
($K=R_{\alpha\beta\gamma\delta}R^{\alpha\beta\gamma\delta}$),
which is the curvature characteristic of the space-time,
also increases sharply along these lines, due to the infinite
compression.

For smaller $u$ the lines $u=const$ go to the Cauchy horizon IAH1
at $v\to\infty$. The increase of the mass function along these
lines is mainly due to the mass inflation and is exponential when
$v\to\infty$. The mass inflation is the basic process in the
formation of the weak singularity instead of the inner smooth
Cauchy horizon.

The weak null singularity shrinks down due to the focusing effect
from the gravity of the $T_{uu}$ radiation and finally comes down to
$r=0$.

\begin{figure*}
\subfigure[ Thick line is position of OAH.  Thin lines are lines of
constant $r$ (values decreasing from bottom to top).
\label{R1a}]{\includegraphics[width=0.48\textwidth]{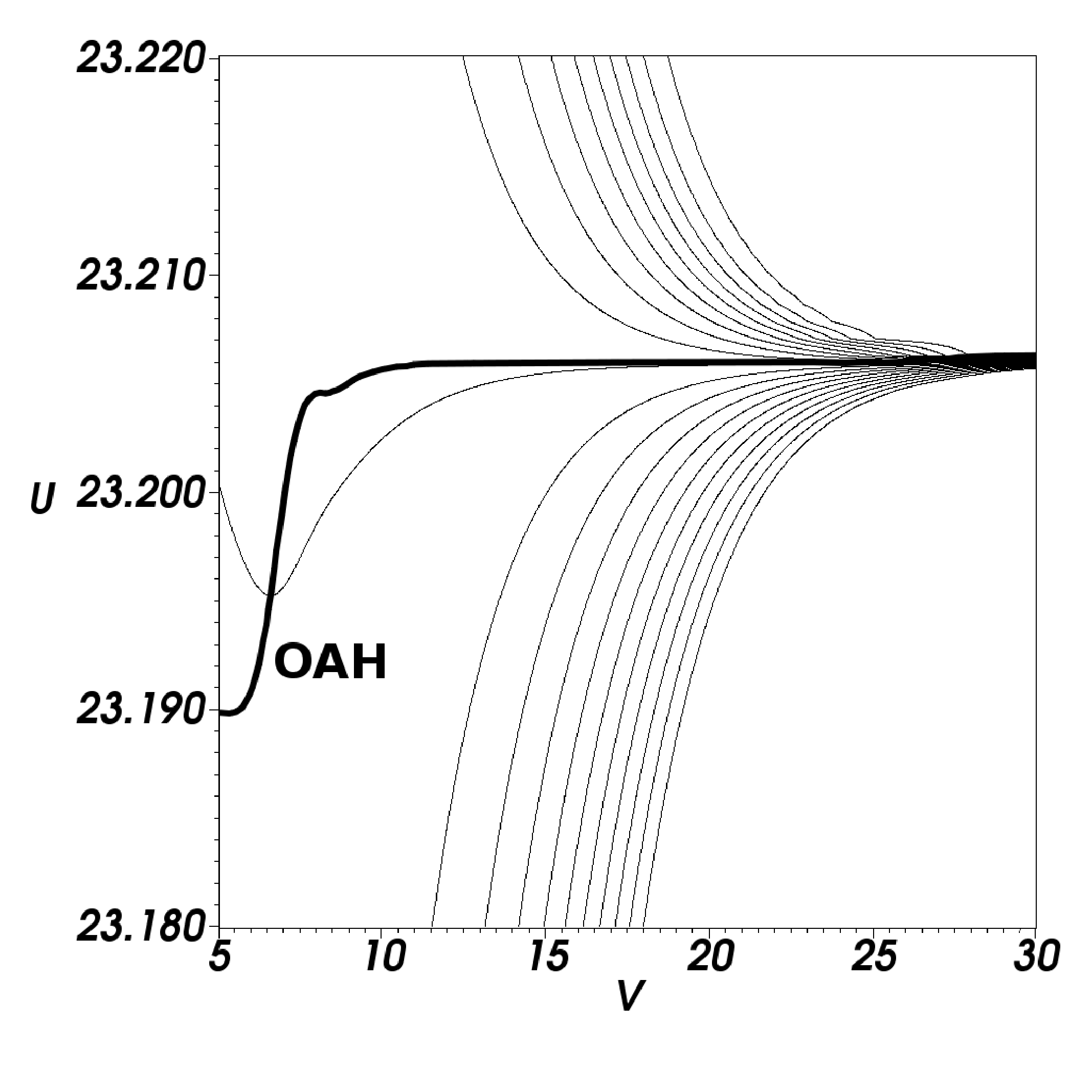}}
\subfigure[Thick line is position of IAH1. Thin lines are lines
of constant $r$ (values decreasing from bottom to top).
\label{R1b}]{\includegraphics[width=0.498\textwidth]{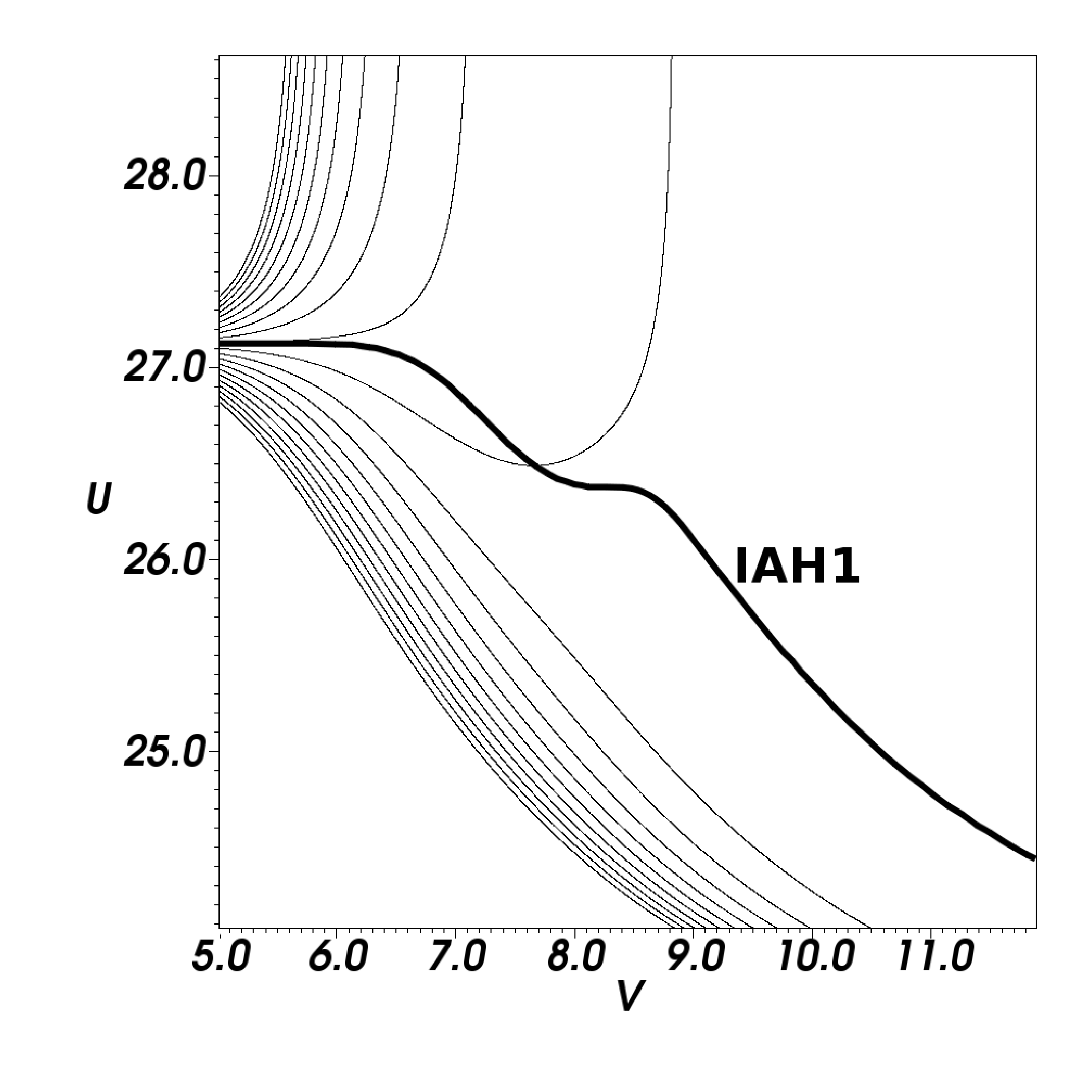}}
\caption{{\label{R1} Positions of the apparent horizons for the
case : $A_\Phi=0$; $A_\Psi=0.01$, ($v_{\Psi 0}=5$, $v_{\Psi
1}=9$). When pulse of the negative energy crosses the outer apparent horizon (OAH), it
becomes smaller and the inner apparent horizon (IAH1) becomes larger.}}
\end{figure*}

\section{Physical processes inside of a charged black hole irradiated by a
pulse of exotic scalar radiation}
\label{sec8}

\begin{figure*}
\subfigure[\label{R2a} Case: $A_\Phi=0$; $A_\Psi=0.01$, ($v_{\Psi
0}=5$, $v_{\Psi 1}=9$).
]{\includegraphics[width=0.48\textwidth]{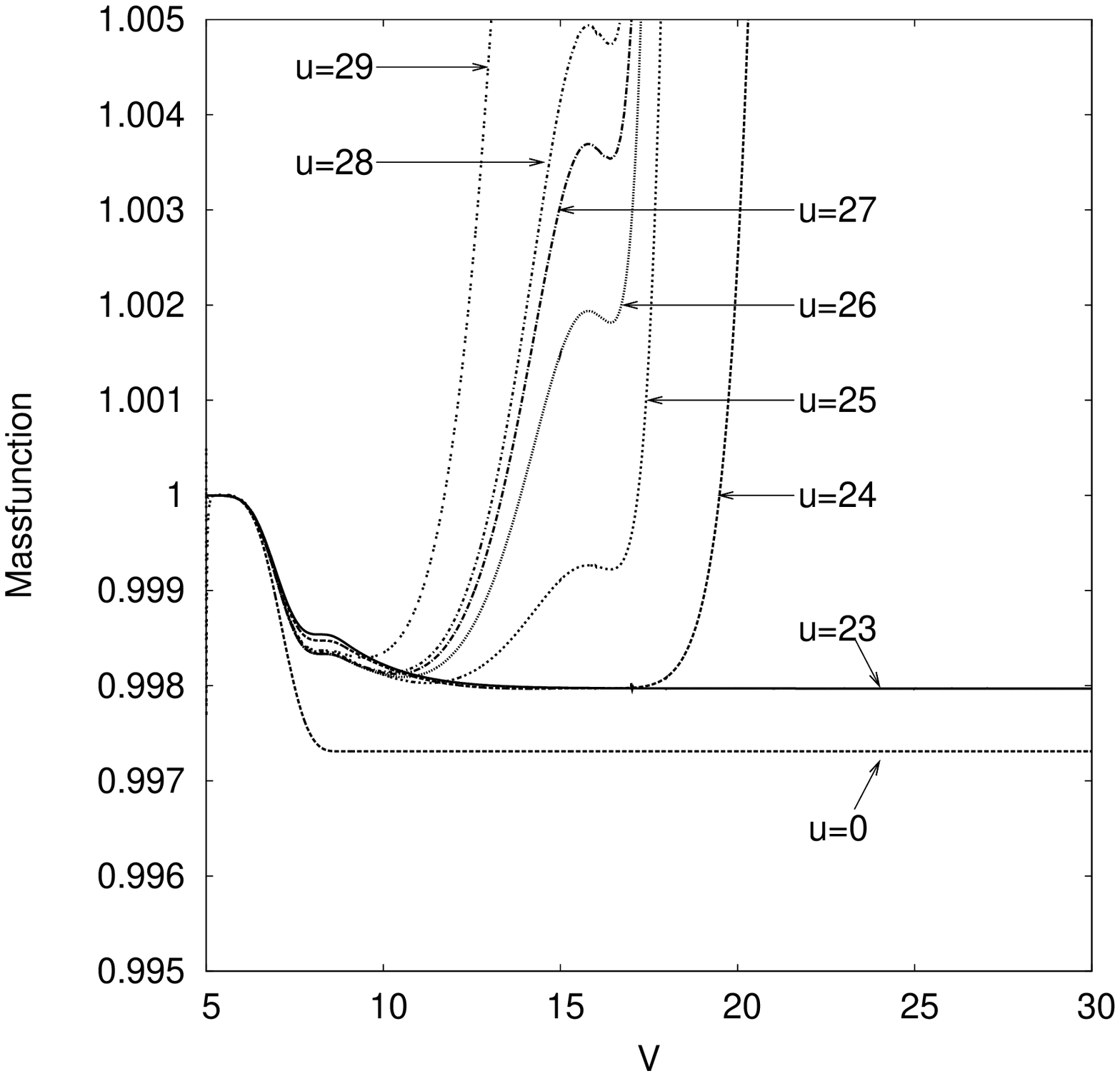}}
\subfigure[\label{R2b} Case: $A_\Phi=0$; $A_\Psi=0.045$, ($v_{\Psi
0}=5$, $v_{\Psi 1}=9$).
]{\includegraphics[width=0.48\textwidth]{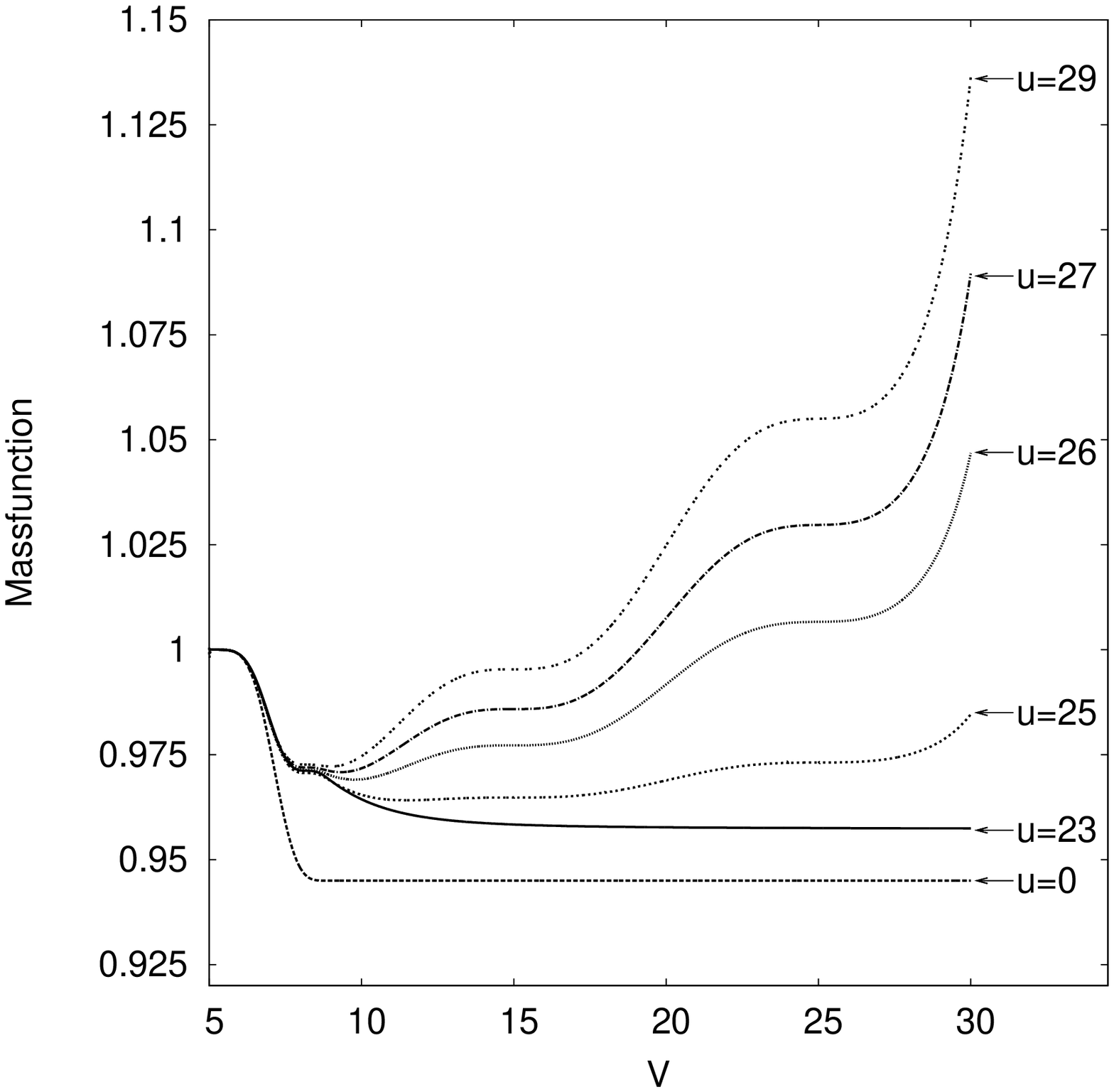}}
\caption{{\label{R2} The mass function along lines of constant $u$
for two cases of weak pulses. One can see the decrease of the mass
function caused by the pulse of negative energy and the
increase of it that results from the mass inflation.}}
\end{figure*}

\begin{figure*}
\subfigure[\label{R3a} Case: $A_\Phi=0$; $A_\Psi=0.01$, ($v_{\Psi
0}=5$, $v_{\Psi 1}=9$).
]{\includegraphics[width=0.48\textwidth]{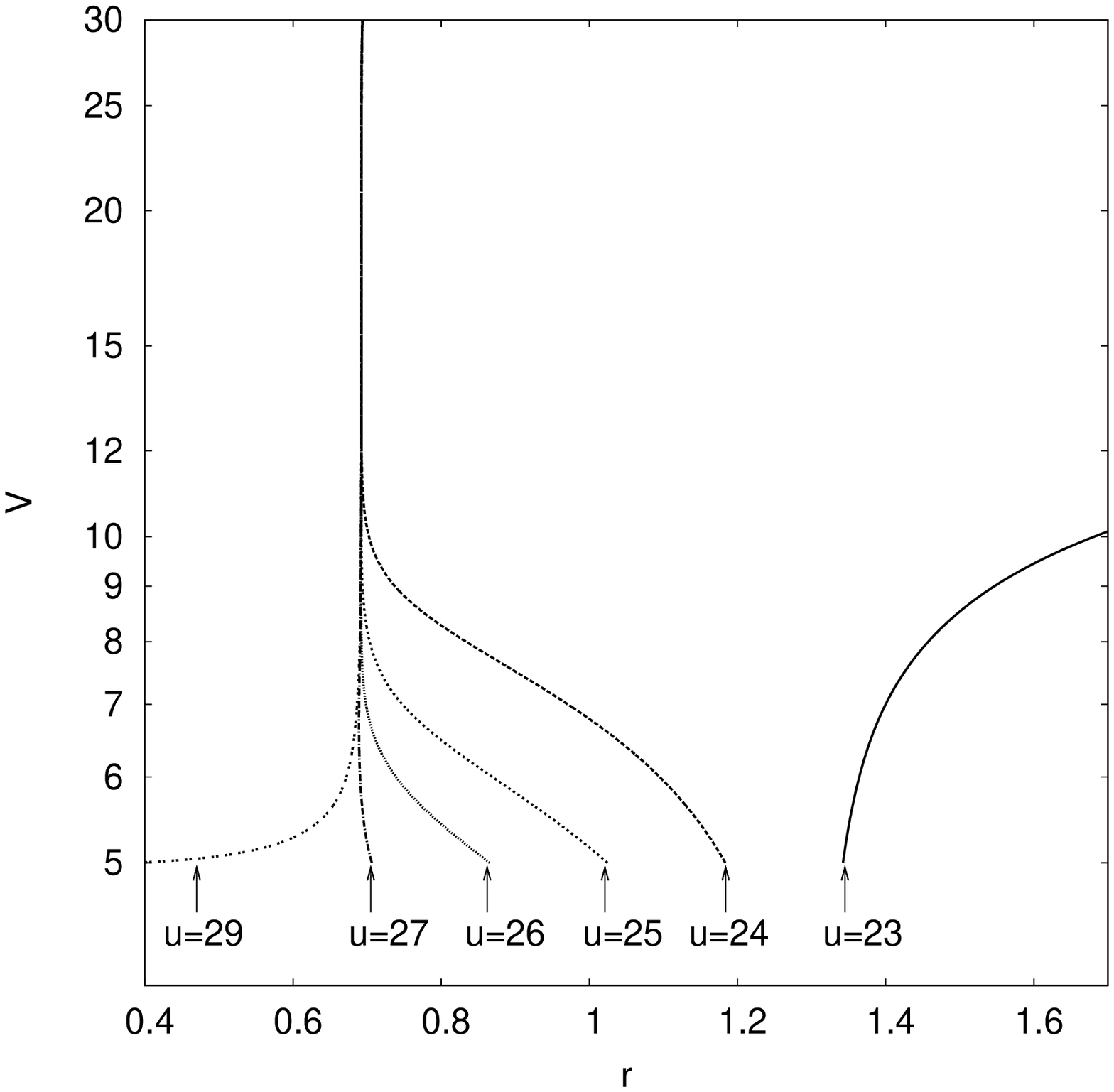}}
\subfigure[\label{R3b} Case: $A_\Phi=0$; $A_\Psi=0.045$, ($v_{\Psi
0}=5$, $v_{\Psi 1}=9$).
]{\includegraphics[width=0.48\textwidth]{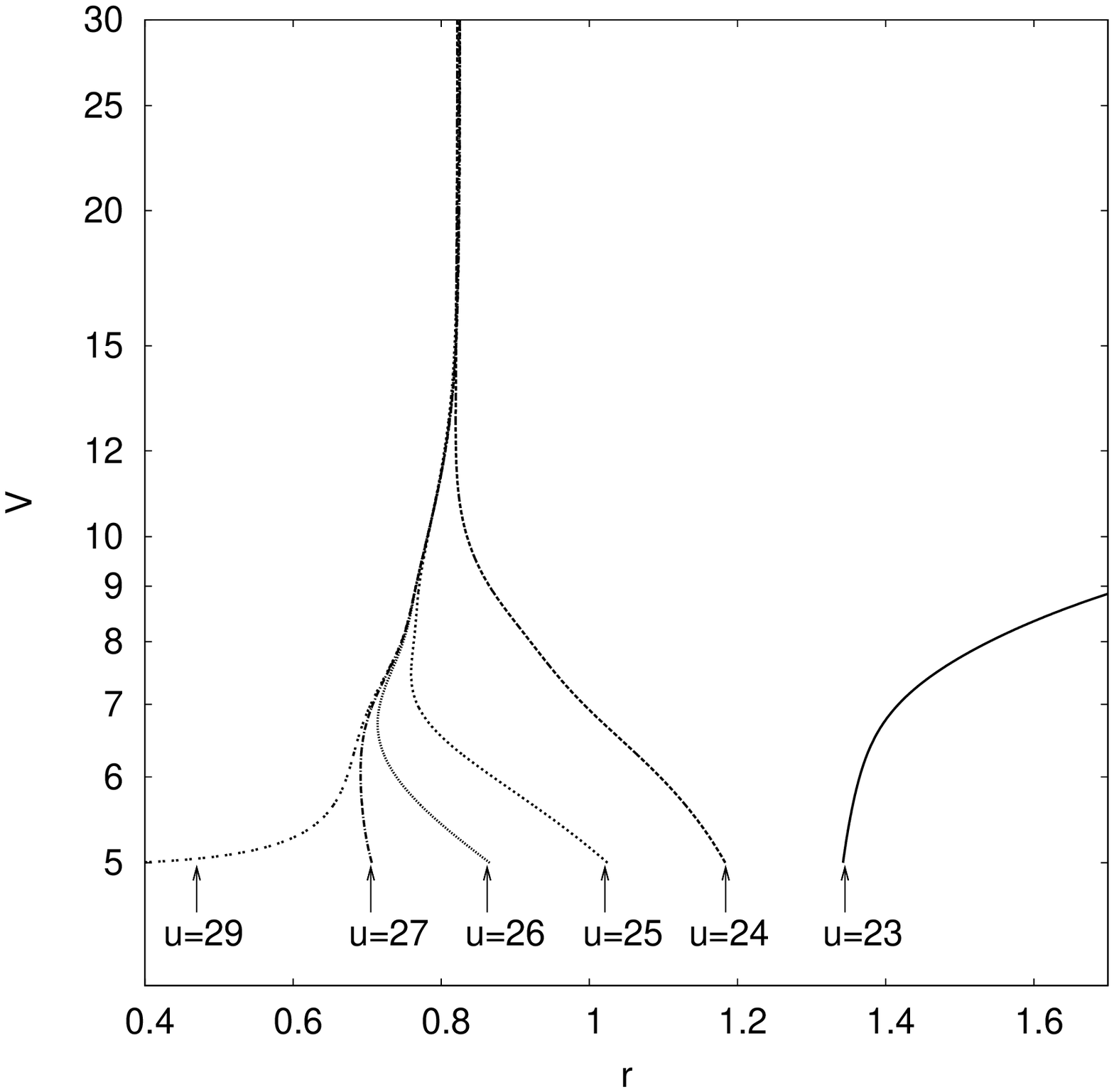}}
\caption{{\label{R3} $v$ versus $r$ along lines of constant $u$
for two cases of weak pulses. All lines with $u\le 24$ comes to
the same $r$ when $v\to\infty$.}}
\end{figure*}

\begin{figure*}
\subfigure[\label{R4a} Case: $A_\Phi=0$; $A_\Psi=0.01$, ($v_{\Psi
0}=5$, $v_{\Psi 1}=9$).
]{\includegraphics[width=0.48\textwidth]{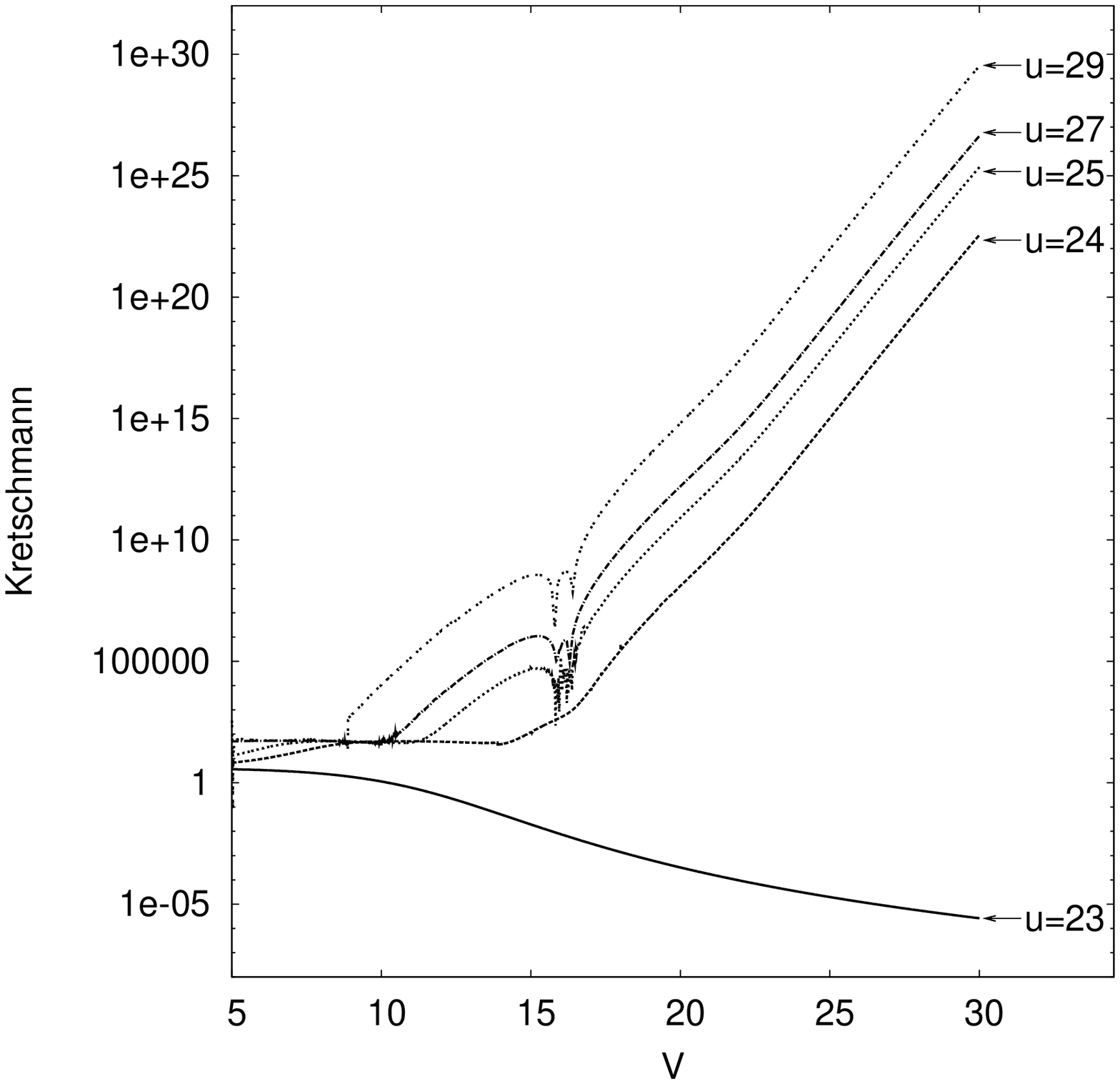}}
\subfigure[\label{R4b} Case: $A_\Phi=0$; $A_\Psi=0.045$, ($v_{\Psi
0}=5$, $v_{\Psi 1}=9$).
]{\includegraphics[width=0.48\textwidth]{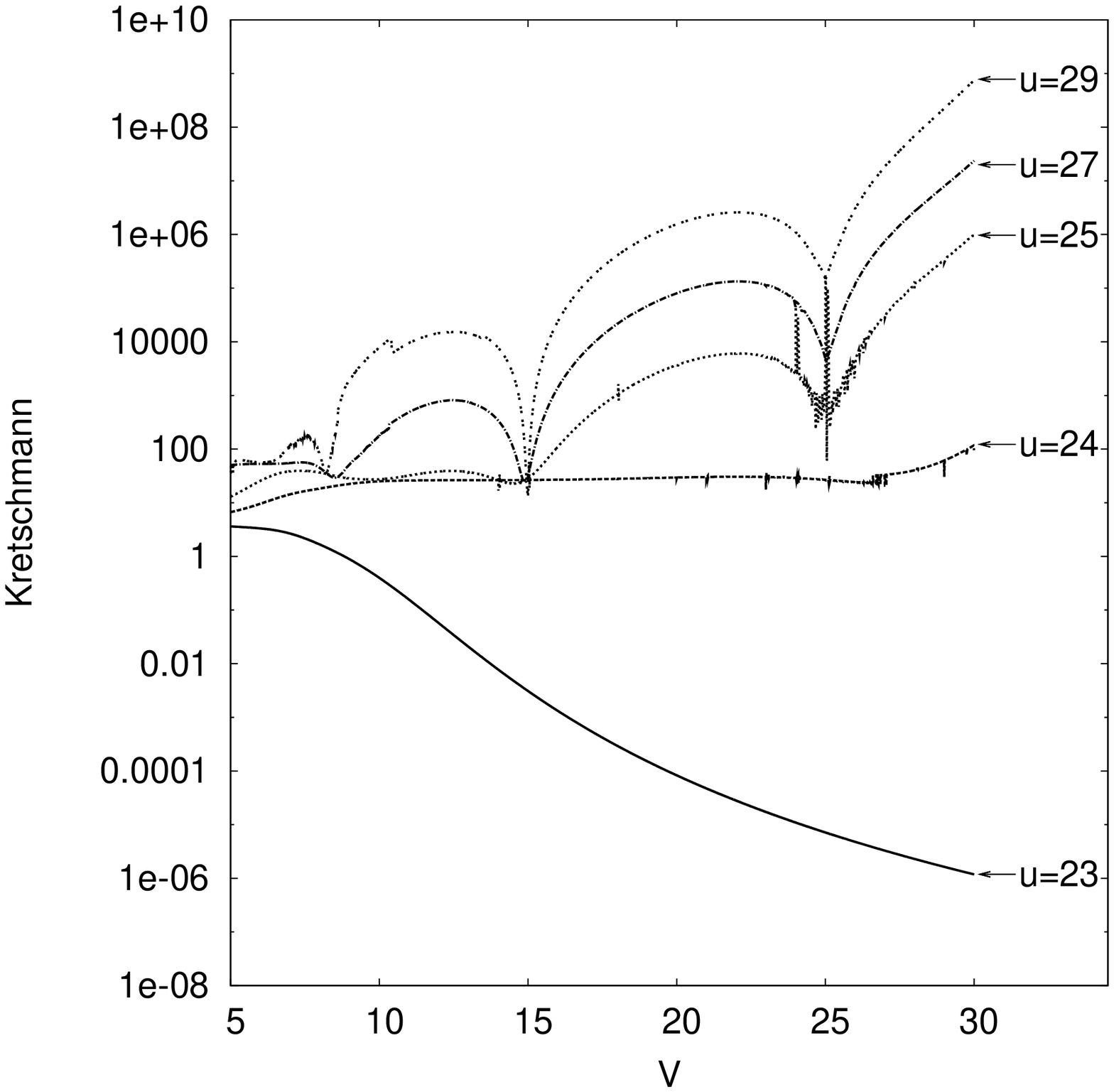}}
\caption{{\label{R4} Kretschmann curvature scalar along lines of constant
$u$ for two cases of weak pulses. All lines inside the BH ($u\le 24$) is seen to increase when $v\to\infty$.}}
\end{figure*}

Now let us compare the processes described in the previous section with the
processes in the case when a Reissner-Nordstr\"om BH is irradiated by
an exotic scalar field.

\begin{figure*}
\subfigure[\label{R5a} Ingoing flux $T_{vv}^{\Psi}$ along lines of
constant $u$.]{\includegraphics[width=0.49\textwidth]{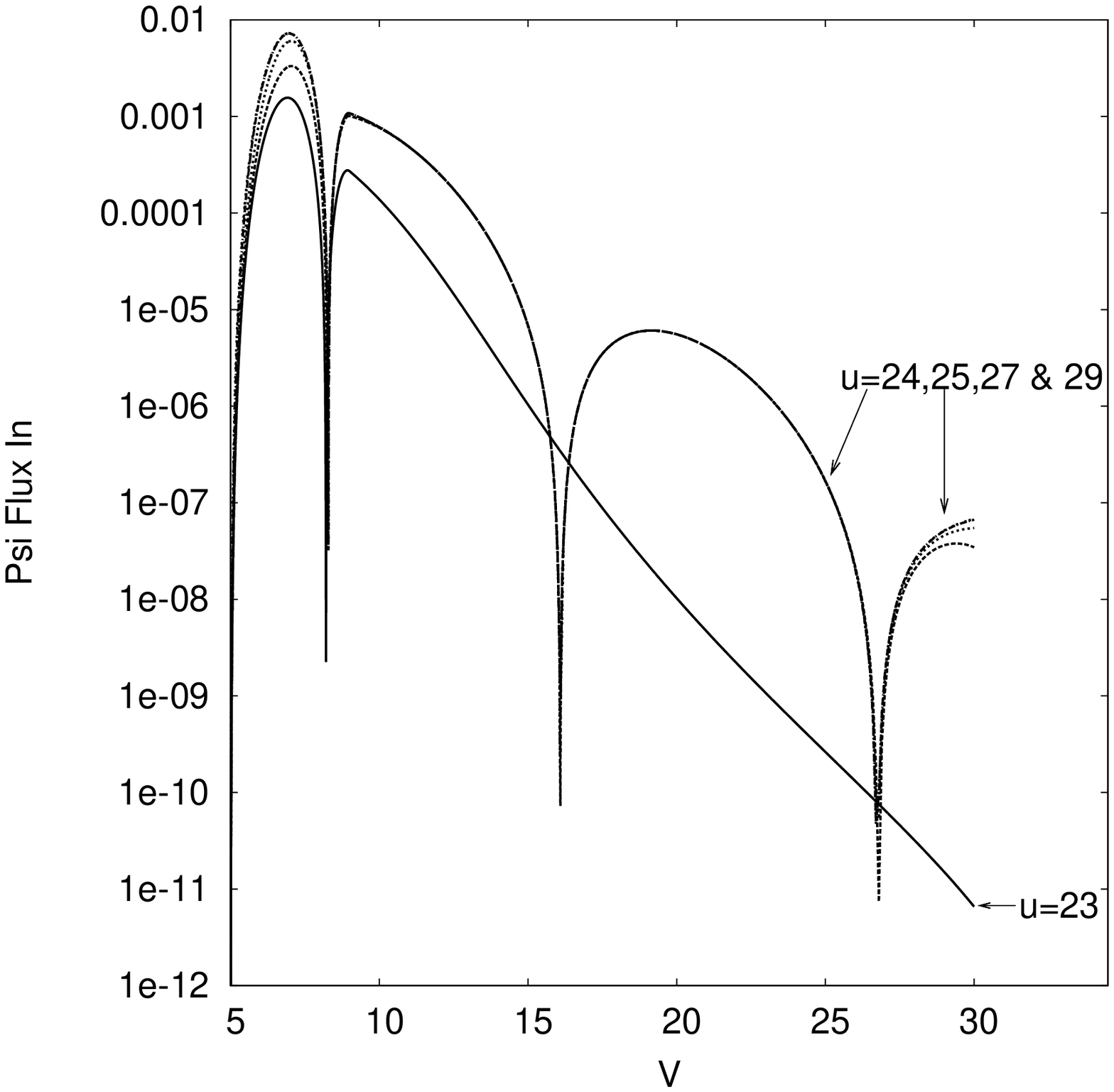}}
\subfigure[\label{R5b} Outgoing flux $T_{uu}^{\Psi}$ along lines
of constant $u$.]{\includegraphics[width=0.49\textwidth]{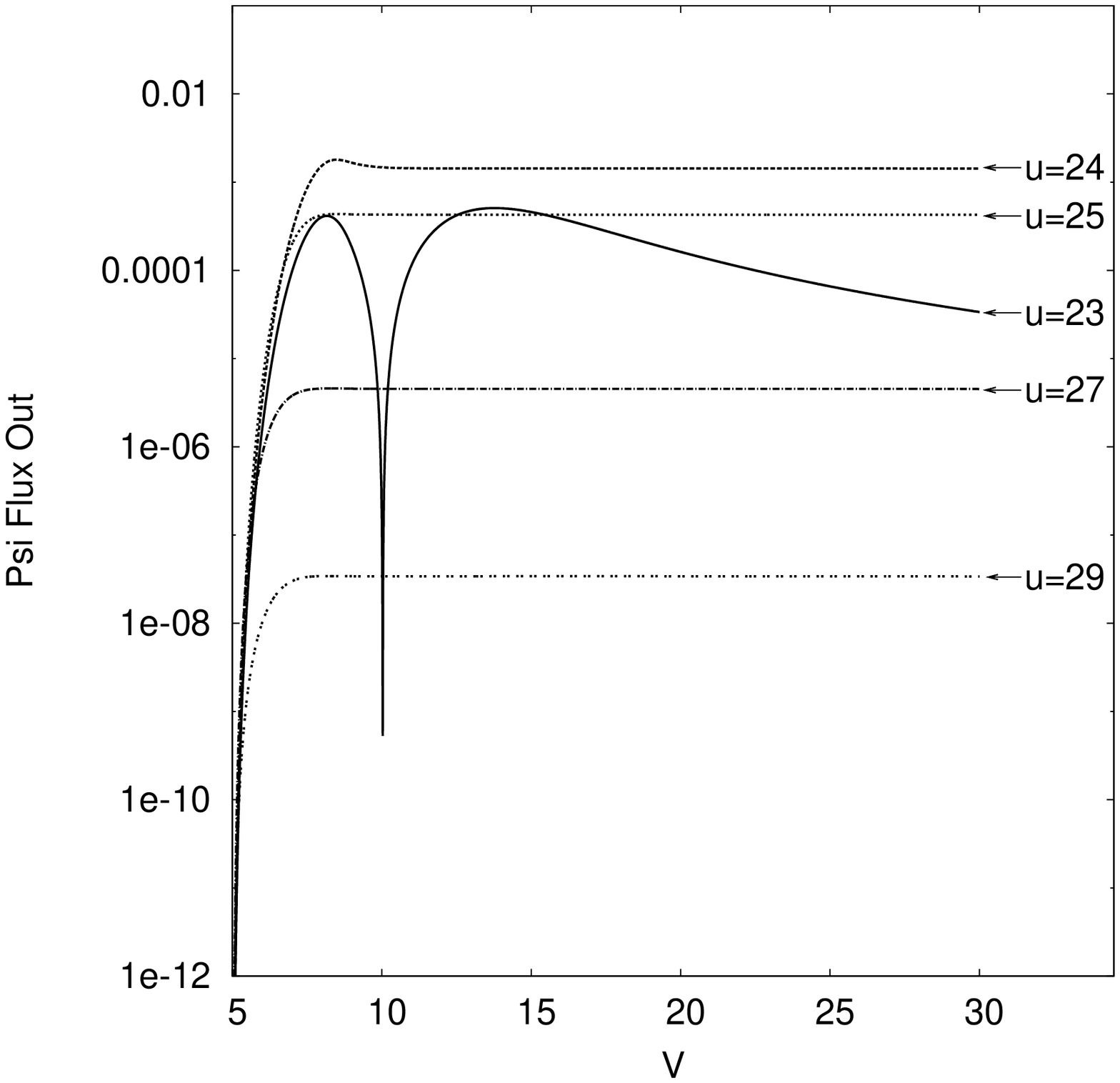}}
\caption{{\label{R5} In- and outgoing $\Psi$ fluxes of the
negative energy for the case: $A_\Phi=0$; $A_\Psi=0.01$ ($v_{\Psi
0}=5$, $v_{\Psi 1}=9$). The initial ingoing pulse can be clearly seen (fig. 5a, between $5\le v\le 9$),
as can the scattered fluxes of the initial pulse.}}
\end{figure*}

\subsection{The case of survival of the BH} \label{sec8-a}
We start from the case when the power of the exotic radiation pulse is rather weak
and the BH survives after being irradiated. Let us first consider the
following case:
\begin{eqnarray}
\label{ex3} A_\Phi=0,;\quad A_\Psi=0.01\quad (v_{\Psi 0}=5,\quad
v_{\Psi 1}=9).
\end{eqnarray}
We will see that total amount of negative mass pumped into the
BH in this case is rather small, it survives but the positions of the apparent horizons
change.

It is clearly seen from fig.~\ref{R1a} and fig.~\ref{R1b} that the
result of the processes with the negative energy density pulse is
opposite to the case of a positive energy density pulse described
above. When the pulse of the negative energy density crosses the
OAH, the horizon becomes smaller and on the corresponding plot
(Fig.~\ref{R1a}) it can be seen to be going to higher $u$
(corresponding to smaller $r$, cf. eq. \eqref{eq:2_10}) in the region $5<v<9$.

Before the pulse at ${v<5}$, the mass of the BH was $m=1$ and the
position of the OAH corresponded to ${u=23.190}$. After the end of the
process, the final position of the OAH
corresponds to ${u\approx 23.206}$ (fig. ~\ref{R1a}) and the final asymptotic external mass is
${m\approx 0.998}$ (as we will see in fig.~\ref{R2a}).

We formulated the initial conditions along the outgoing null
surface ${u=0}$. When the ingoing radiation pulse propagates from
${u=0}$ to the OAH, it experiences scattering because of the
curvature of the space-time, thus being converted into outgoing
radiation. At later times, that scattered radiation is then
rescattered and being converted back into ingoing radiation and
so on and so forth. The tail of this scattered radiation thus
reaches the OAH at later times. This can be observed in fig.
~\ref{R1a}) where the OAH can be seen to change its position in
two steps.

We would like to emphasize that the scattering and the
rescattering are the result of interaction of the propagating
field with the curvature of the space-time background but not any
boundary conditions.

The first step, when the initial pulse passes into the BH, can be
observed between ${5 \lesssim v \lesssim 9}$, and the second step
(caused by the scattered radiation) can be observed between ${9
\lesssim v \lesssim 11}$.

When the pulse crosses the IAH1, the effects are again opposite to those of an energy pulse with a positive energy
density and the horizon thus becomes bigger (goes
to smaller $u$) in the region $5<v<9$ (Fig.~\ref{R1b}).

In the region ${v \gtrsim 9}$ in Fig.~\ref{R1b}, the internal
horizon IAH1 is seen to go to smaller $u$ (higher $r$). This is
the manifestation of the antifocusing gravitational effects which
are opposite to the effect 2), mentioned above in the discussion
of the positive energy density pulse (see section~\ref{sec7}).

The behavior of the mass function inside the pulse in the region
$5 \lesssim v \lesssim 9$, (shown in Fig.~\ref{R2a} along lines of constant $u$), is also
opposite to the case of an ingoing pulse with positive energy
density described in section \ref{sec7}, in that the mass function decreases as the pulse enters the BH.
This is also the case for pulses
with higher (negative) energy contents (Fig.~\ref{R2b}).
In Fig.~\ref{R2a} it can be seen that the mass function along the line $u=0$ (where we formulate the initial conditions) is lower than
along the line $u=23$ (slightly outside of the OAH). This is the manifestation of the scattering processes outside the BH.
Because of the scattering process, only part of the energy along $u=0$ will actually reach the OAH near $u=23$, the rest will be converted
into outgoing radiation that will forever escape from the BH.

The consequence of the concentration of the incoming energy near
the inner Cauchy horizon, is still the same as it was in the case
of the positive energy density pulse. This concentration can be
seen in Fig.~\ref{R3a} where lines of $r$ versus $v$ along
$u=const$ asymptotically tend to $r=r_{Cauchy}$ for $u$ big
enough.

The sharp increase of the mass-function for the greater $v$ inside
the BH (for ${v\gtrsim 12}$) in Fig.~\ref{R2a} is the result of
the mass inflation which still works in the case of an exotic radiation
pulse.

The Kretschmann scalar along $u=const$ is shown in
Fig.~\ref{R4a}. It is seen to increase with $v$ for high values of $u=const$.
The exponential increase of the Kretschmann scalar with $v$ is a
fingerprint of coming to the weak singularity at ${v=\infty}$.
Thus in this case, probably the weak singularity exists at the
Cauchy horizon.

Fig.~\ref{R5a} represents the evolution of the in-flux, $T_{vv}$,
of the negative scalar energy into the BH. Fig.~\ref{R5b} shows the corresponding $T_{uu}$
out-flux which arises as a result of the $T_{vv}$ flux being
scattered by the space-time curvature. The tails of the $T_{vv}$
flux are the result of backscattering of the $T_{uu}$ flux and so
on and so forth.

In both Fig.~\ref{R5a} and Fig.~\ref{R5b} one can see the
resonances arising as a result of the scattering process. These
resonances are typical also for the cases of the irradiation by
the more intense pulses.

Now let us consider the case of a more intense pulse of negative
energy, but still not strong enough to destroy the BH.
We consider the case
\begin{eqnarray}
\label{ex5-045} A_\Phi=0;\quad A_\Psi=0.045, \quad (v_{\Psi
0}=5,\quad v_{\Psi 1}=9)
\end{eqnarray}

\begin{figure}
\includegraphics[width=0.48\textwidth]{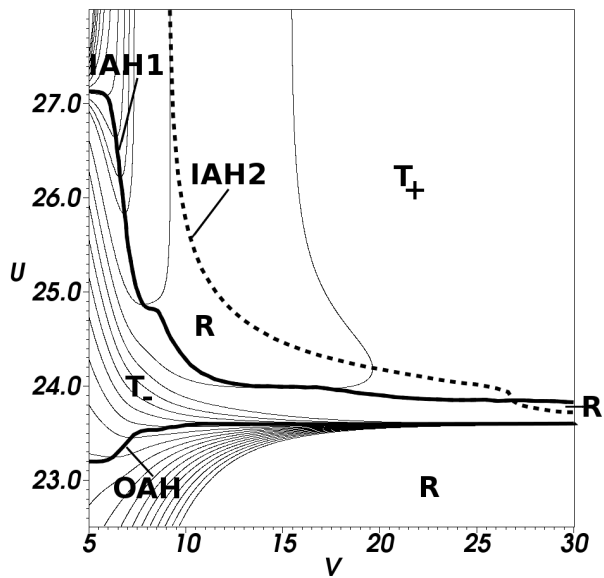}
\caption{{\label{R6} Lines of constant $r$ (thin lines, decreasing from bottom to top),
positions of $R$- and $T$-regions and
positions of the apparent horizons (thick lines) for the case of $A_\Phi=0$;
$A_\Psi=0.045$, ($v_{\Psi 0}=5$, $v_{\Psi 1}=9$). In this case, the BH survives as can be seen by the OAH
and IAH1 not meeting.}}
\end{figure}

First of all we note the following; In Fig.~\ref{R2b} one can see
that at ${u=0}$ the value of the mass function after the pulse
${v>9}$ corresponds to the external mass ${m\approx 0.945}$. This
is less than ${q=0.95}$ which means, that if this value does not
change when one comes to the OAH, the BH must be destroyed. But
because of the scattering process, part of the negative energy
of the pulse will be scattered away during the propagation of the
pulse from ${u=0}$ to the OAH and the amount of the negative energy
pumped into the BH will be smaller. As a consequence, the final mass
of the BH after the end of the whole process will be ${m\approx
0.959}$, and the position of the OAH process will be ${u\approx
23.6}$, see Fig.~\ref{R6}. Thus the BH survives.

The reason for the increase of the mass function (in
Fig.~\ref{R2b}) in the region just after ${v\approx 10}$ for great
$u$ is mainly the additional flux of the negative energy $T_{uu}$
which arise because of the scattering effect (analogous to
the scattering demonstrated in Fig.~\ref{R5}) and for bigger $v$ it is a mild mass inflation.

The properties of the Kretschmann scalar inside of the BH is now
more complicated. At big $u$ it shows growing oscillations with
$v$, see Fig.~\ref{R4b}. These oscillations are the consequence of the oscillations of the
$T_{uu}$ and $T_{vv}$ fluxes which are the analogues to those in
Fig.~\ref{R5}.

Finally in Fig.~\ref{R6} one can see the general structure of the $R$-
and $T$-regions. For all situations we will call the OAH the
horizon which is practically the border of the BH. The IAH1 is the
border between the $T_-$-region and the internal $R$-region and
its continuation. The horizon IAH2 is the border between the
internal $R$ and $T_+$ regions and its continuation. All these
horizons can be seen in Fig.~\ref{R6}. Most interesting is the
fact that the border IAH2 is not along a null-geodesic line
(Cauchy Horizon) but comes to be visible inside the domain of
integration.

\begin{figure}
\includegraphics[width=0.48\textwidth]{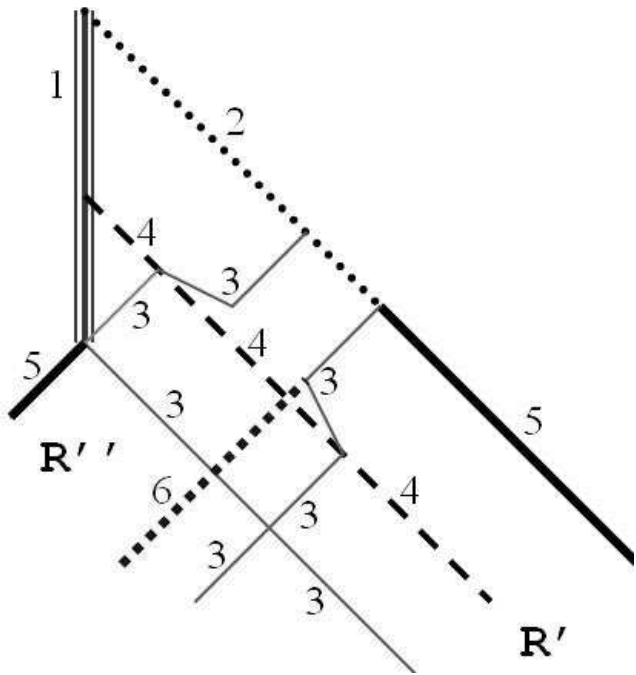}
\caption{{\label{R7} The Penrose diagram for the case when the BH
survives after being irradiated by a pulse of the exotic scalar
field. Here: 1 -- strong, $r=0$, timelike singularity, 2 -- weak
singularity, 3 -- horizons, 4 -- narrow pulse of the exotic scalar
field, 5 -- light infinity, $R'$ -- our universe, $R''$ -- another
universe, 6 -- the event horizon.}}
\end{figure}

On the basis of our computations and their analysis we may guess at
the Penrose diagram in Fig.~\ref{R7} for the case when the BH
survives after being irradiated by a pulse of the exotic scalar field.

\subsection{Destruction of the BH}
\label{sec8-b}
\begin{figure}
\subfigure[\label{R8a} Case: $A_\Phi=0$; $A_\Psi=0.05045$.
]{\includegraphics[width=0.45\textwidth,height=0.38\textwidth]{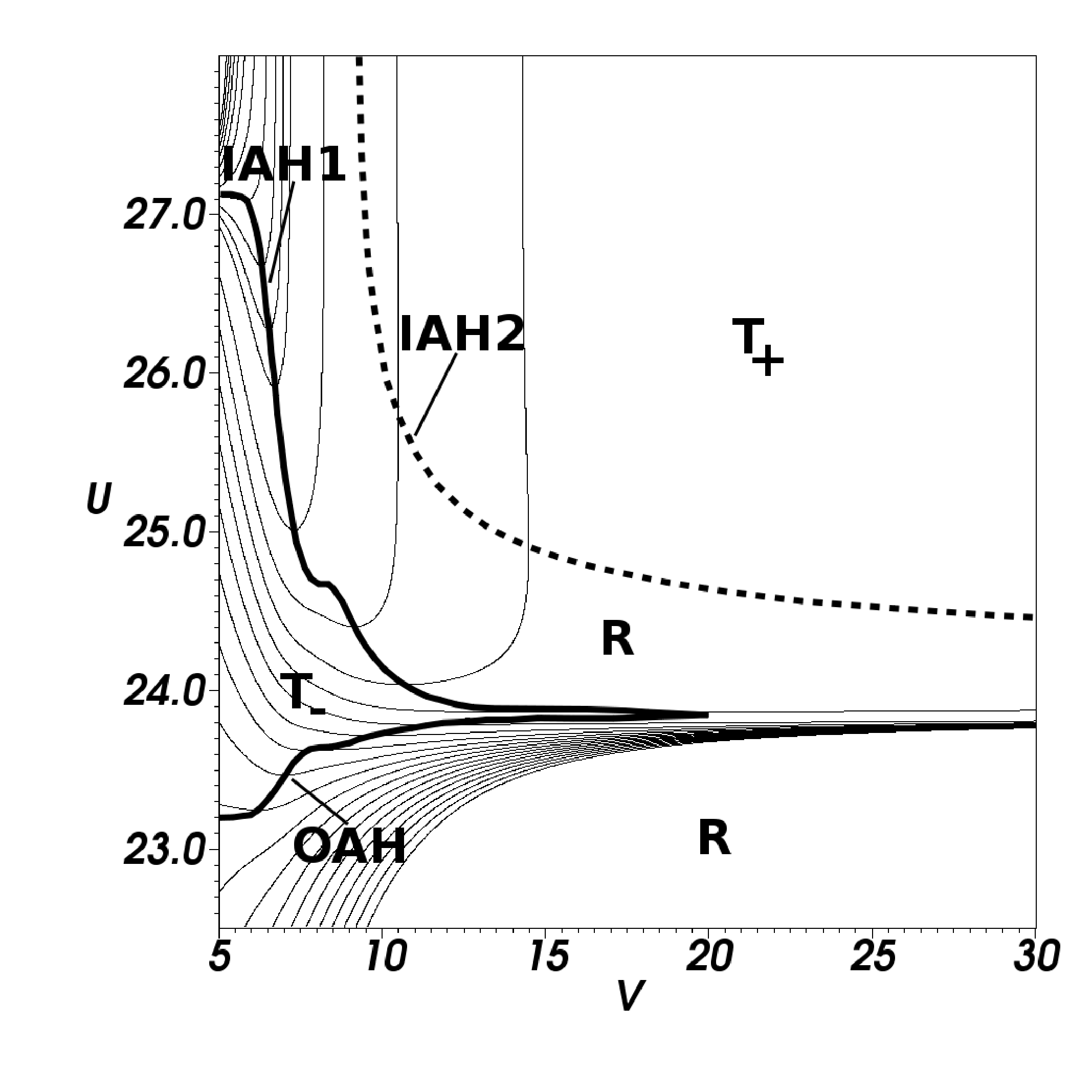}}
\subfigure[\label{R8b} Case: $A_\Phi=0$; $A_\Psi=0.0513$.
]{\includegraphics[width=0.45\textwidth,height=0.38\textwidth]{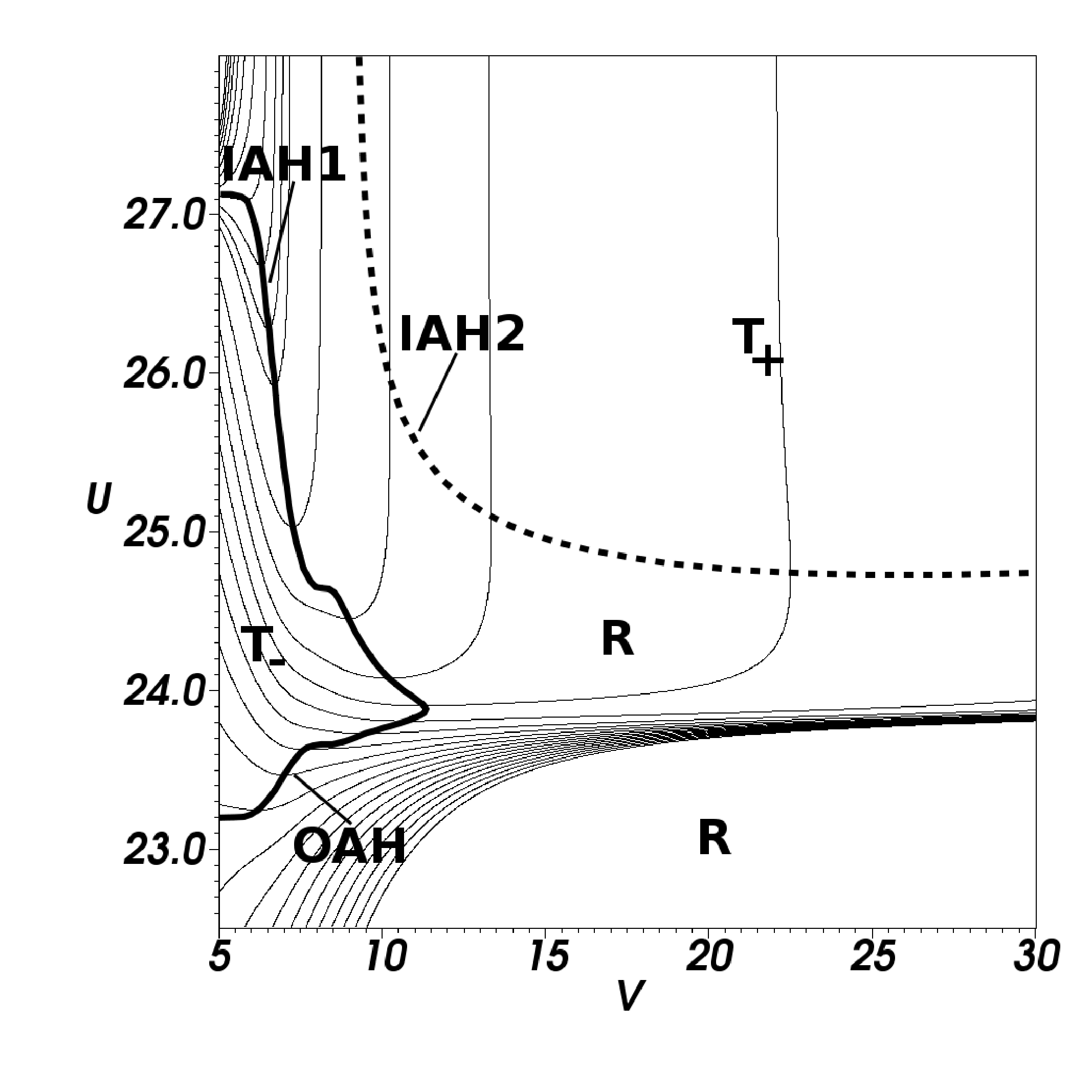}}
\subfigure[\label{R8c} Case: $A_\Phi=0$; $A_\Psi=0.070$.
]{\includegraphics[width=0.45\textwidth,height=0.38\textwidth]{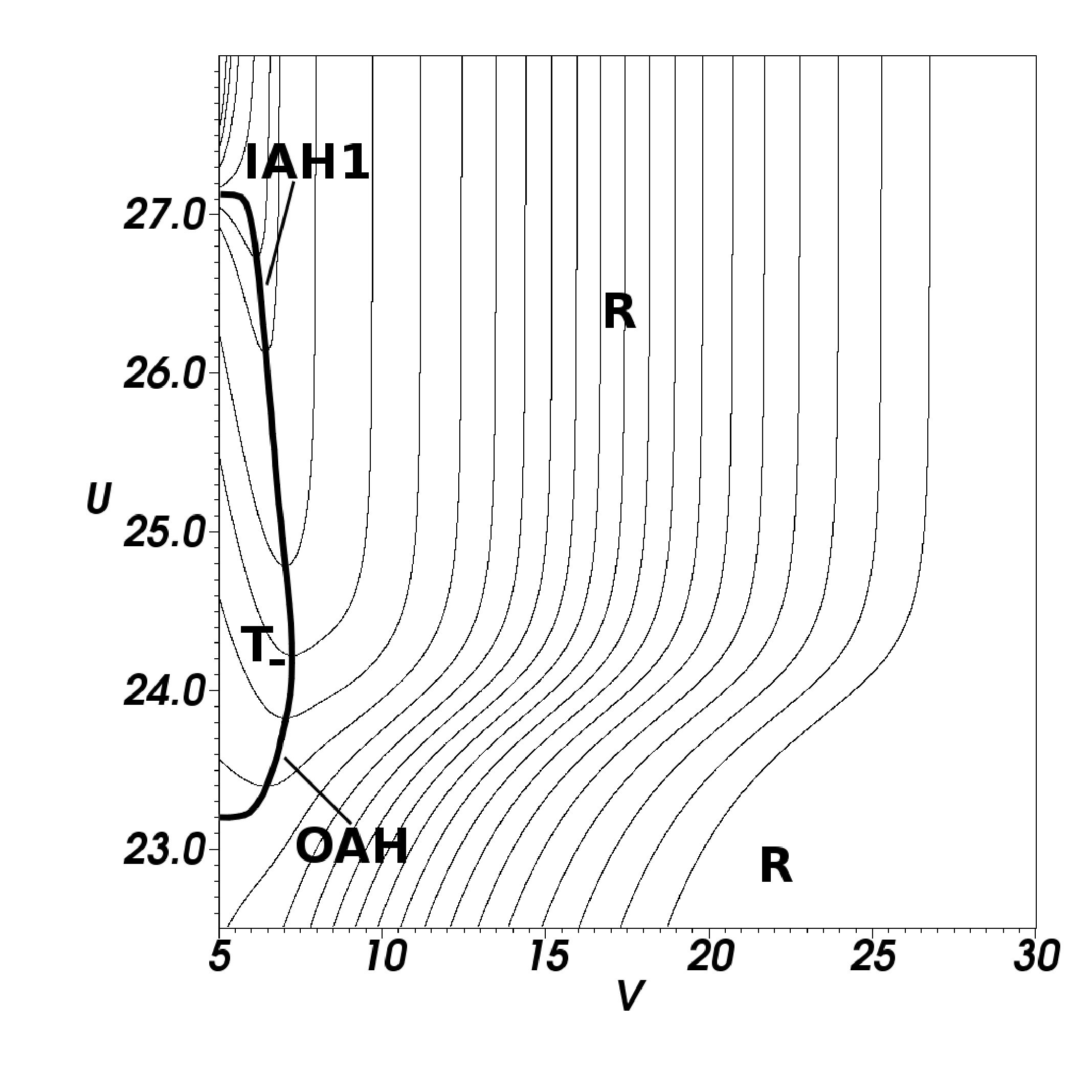}}
\caption{{\label{R8} Lines of constant $r$ (thin lines, decreasing from bottom to top), positions of $R$- and $T$-regions and
positions of the apparent horizons (thick lines) for the cases a) $A_\Psi=0.05045$; b) $A_\Psi=0.0513$; c)
$A_\Psi=0.070$. In all cases ($v_{\Psi 0}=5$, $v_{\Psi 1}=9$) and
in all cases the BH is destroyed as can be clearly seen by the meeting of the horizons.}}
\end{figure}

Next we consider the following cases with (higher) negative energies of the
exotic pulse:
\begin{eqnarray}
\label{ex5-a} A.\quad A_\Phi=0;\quad A_\Psi=0.05045 \\
\label{ex5-b} B.\quad A_\Phi=0;\quad A_\Psi=0.0513 \\
\label{ex5-c} C.\quad A_\Phi=0;\quad A_\Psi=0.0700
\end{eqnarray}
In all cases ${v_{\Psi 0}=5}$, ${v_{\Psi 1}=9}$.
Figs.~\ref{R8a}-\ref{R8c} show the further evolutions of the
general picture of the $R$ and $T$ regions for these cases.

\begin{figure}
\subfigure[\label{R9a} $A_\Psi=0.05045$
]{\includegraphics[width=0.45\textwidth,height=0.38\textwidth]{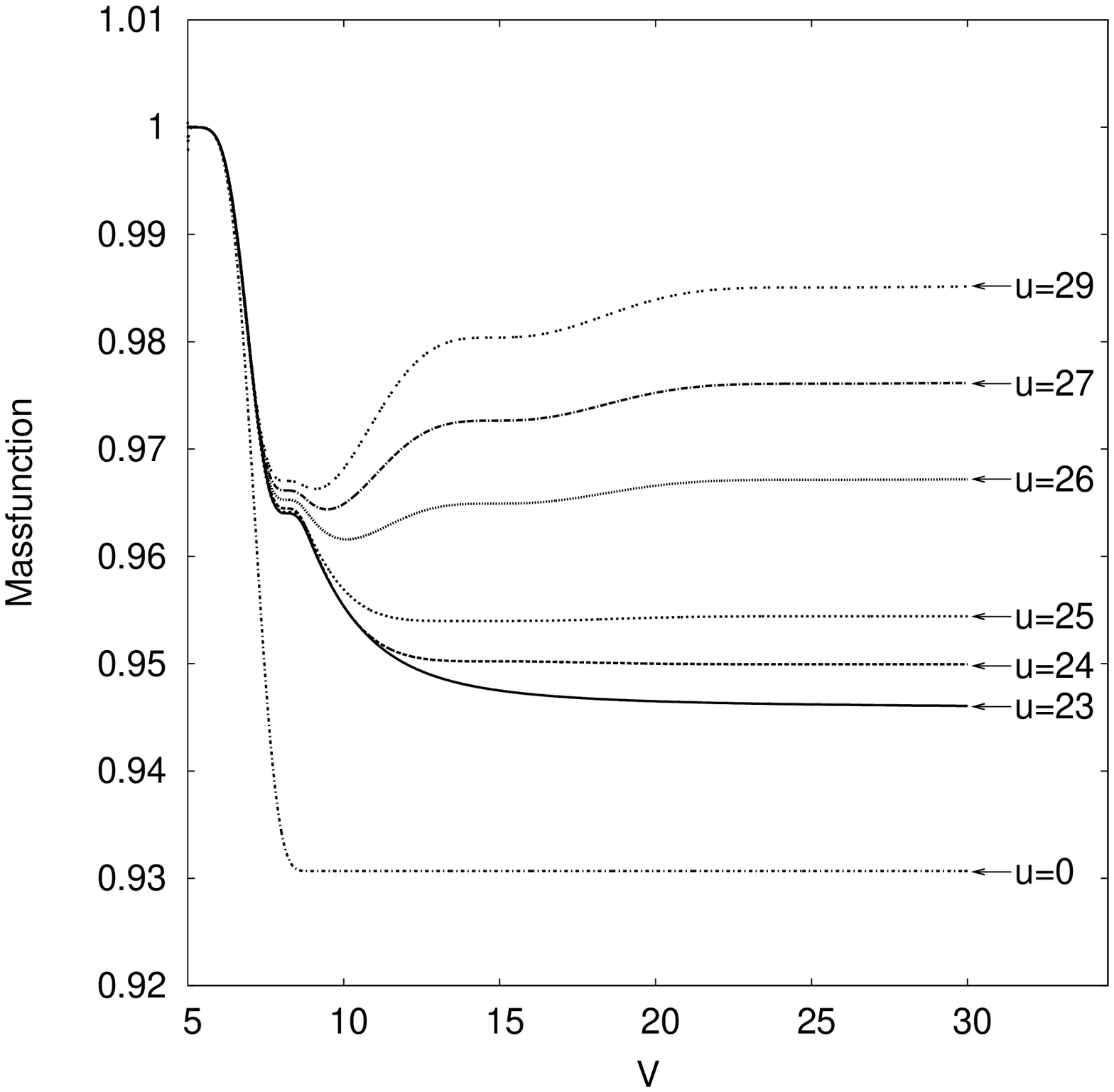}}
\subfigure[\label{R9b} $A_\Psi=0.0513$
]{\includegraphics[width=0.45\textwidth,height=0.38\textwidth]{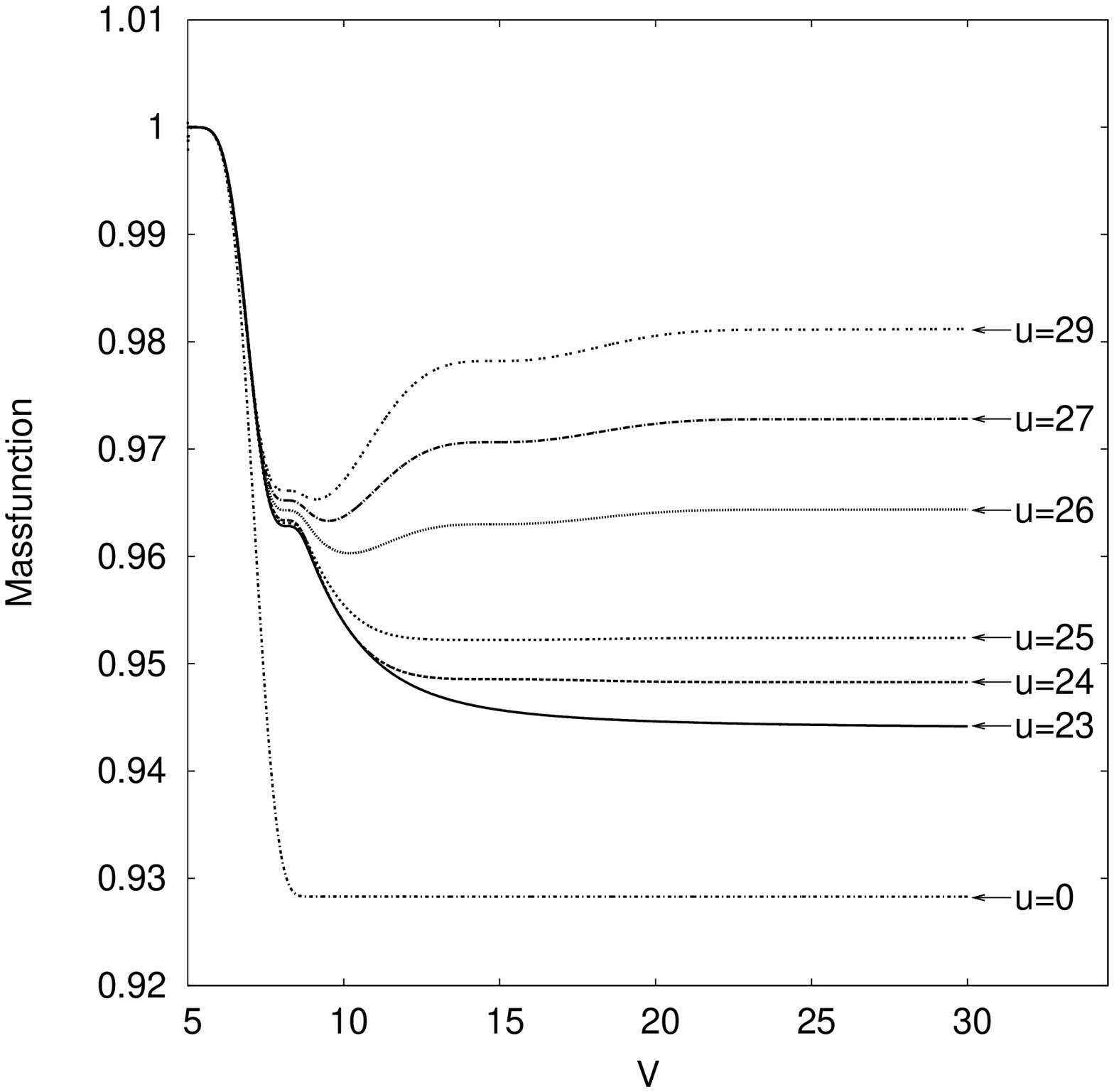}}
\subfigure[\label{R9c} $A_\Psi=0.070$
]{\includegraphics[width=0.45\textwidth,height=0.38\textwidth]{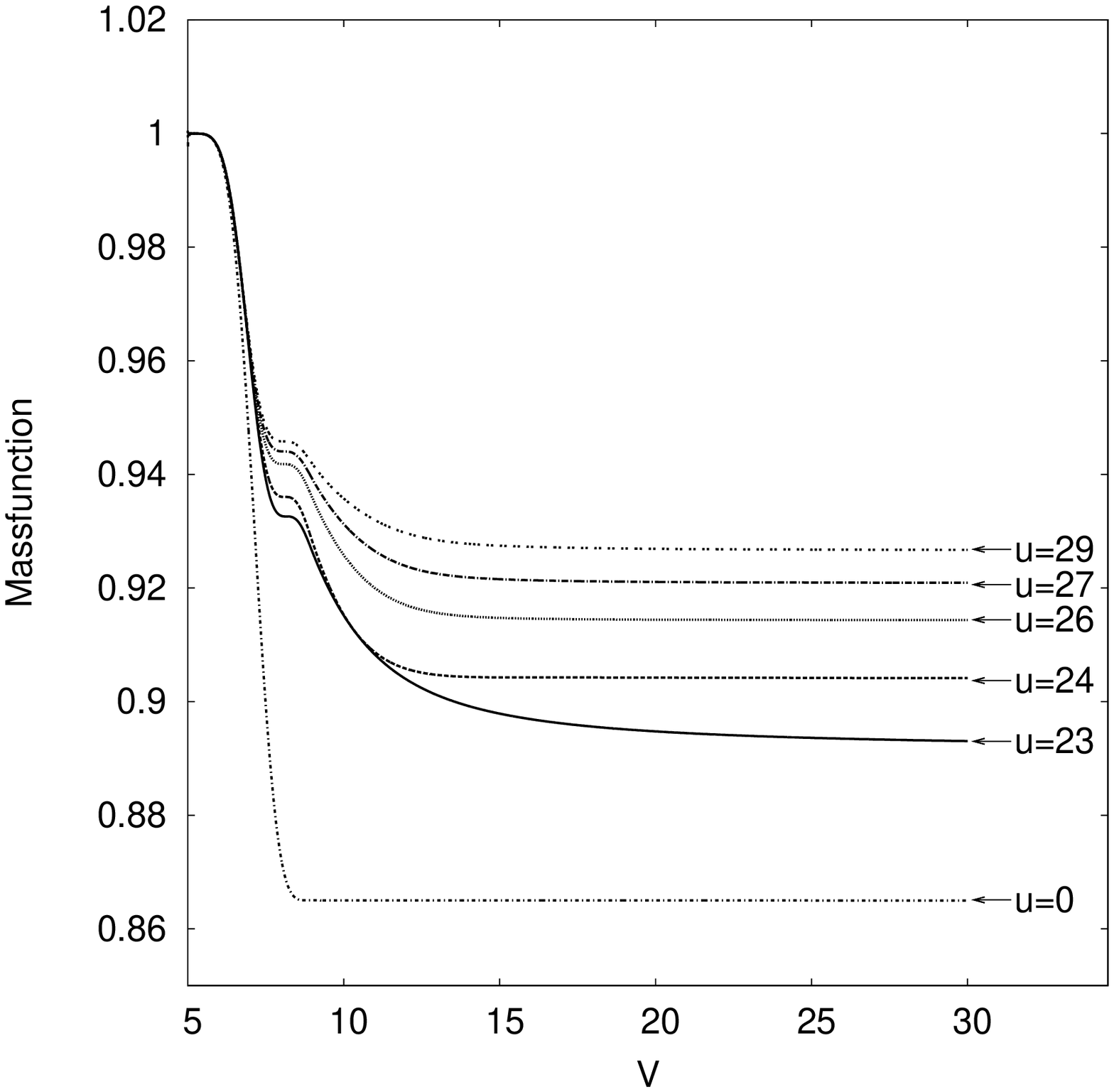}}
\caption{{\label{R9} The mass function along lines of constant $u$
for the cases: a) $A_\Psi=0.05045$; b) $A_\Psi=0.0513$; c)
$A_\Psi=0.070$. In all cases $A_\Phi=0$; ($v_{\Psi 0}=5$, $v_{\Psi
1}=9$) and in all cases the BH is destroyed. }}
\end{figure}

In all these cases, the
total power of the exotic pulse is large enough to reduce the mass
of the object in the critical region ${u\approx [23.2; 23.9]}$ (
where the OAH formed for the previous cases), to below the critical value
${m_{crit}=q=0.95}$, see Fig.~\ref{R9}. Thus even after the
reduction of the power of the pulse during the propagation from
${u=0}$ to ${u\approx 23.2}$ due to the scattering, it is
strong enough to destroy the BH.

This means that the outer and inner apparent horizons should meet and
disappear. This process is seen in Fig.~\ref{R8a} and even more
clearly in Fig.~\ref{R8b}, which shows the slightly stronger
case~(\ref{ex5-b}).

For these cases, for high $v$, the BH is converted into an object
where the outer $R$ and inner $R$-regions are connected. Now the
test photons $u=const$ for all value of $u$ go to bigger $r$, when
$v\to\infty$.

\begin{figure}
\subfigure[\label{R10a} Case: $A_\Psi=0.05045$
]{\includegraphics[width=0.45\textwidth,height=0.38\textwidth]{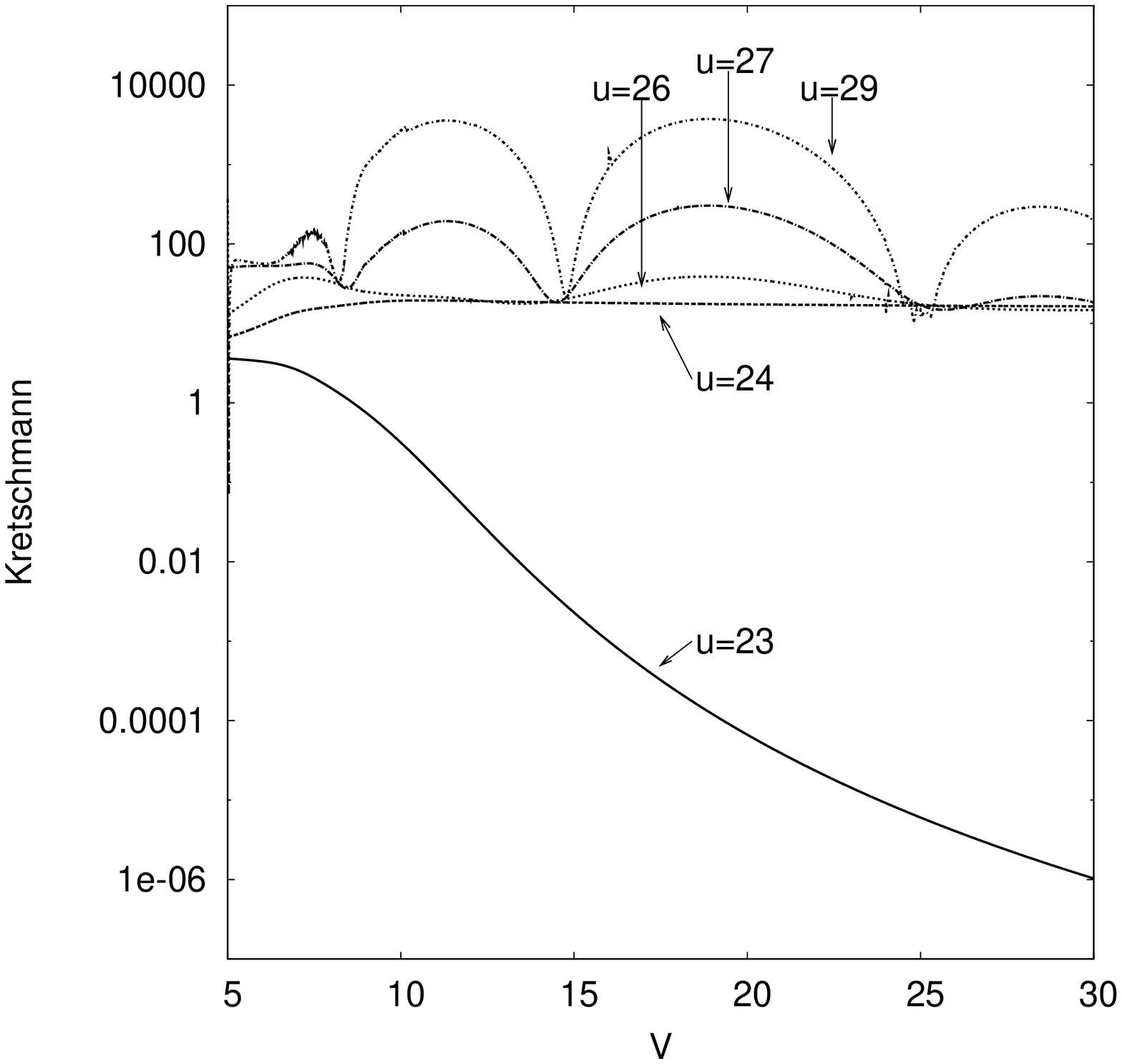}}
\subfigure[\label{R10b} Case: $A_\Psi=0.0513$
]{\includegraphics[width=0.45\textwidth,height=0.38\textwidth]{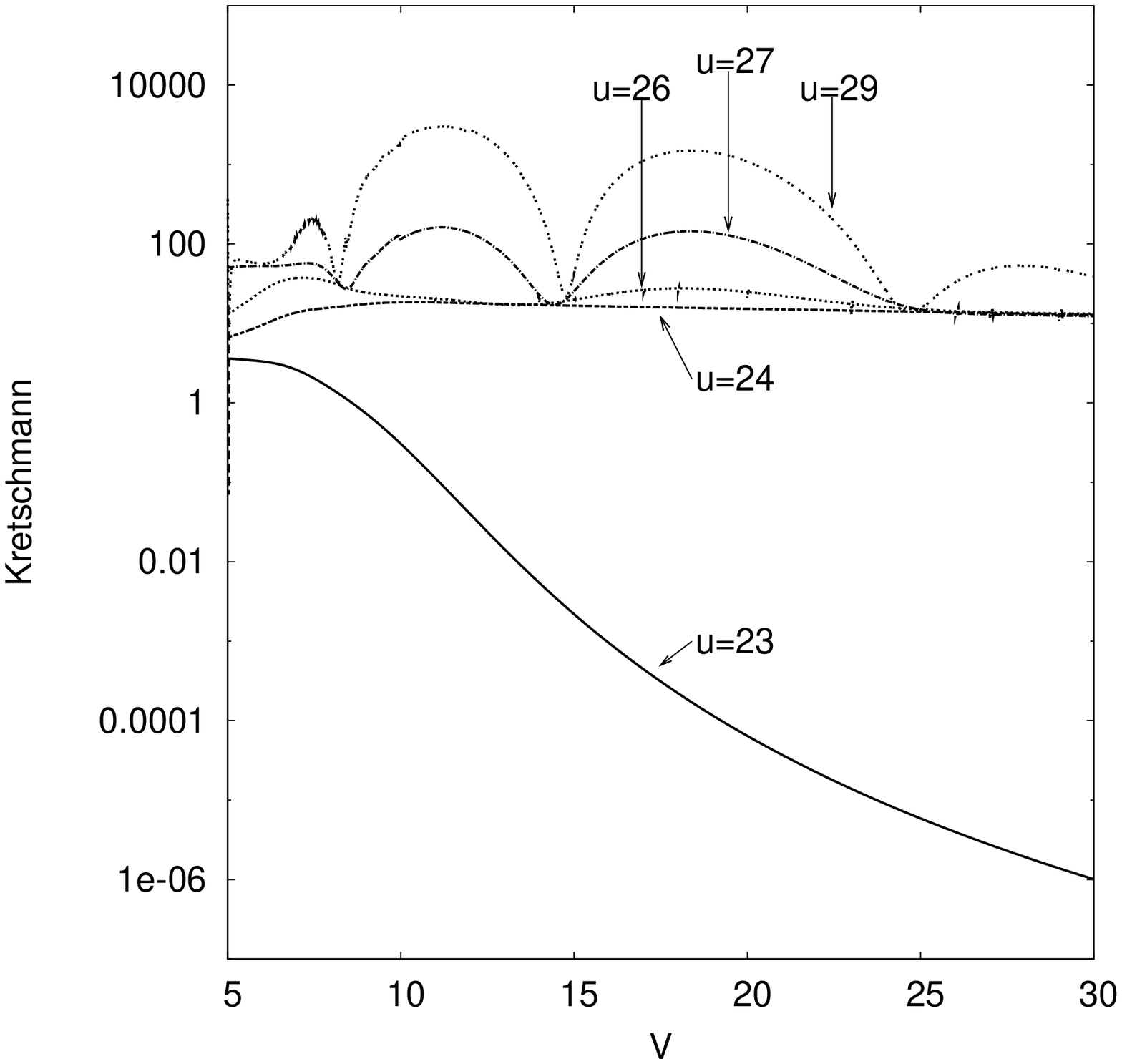}}
\subfigure[\label{R10c} Case: $A_\Psi=0.070$
]{\includegraphics[width=0.45\textwidth,height=0.38\textwidth]{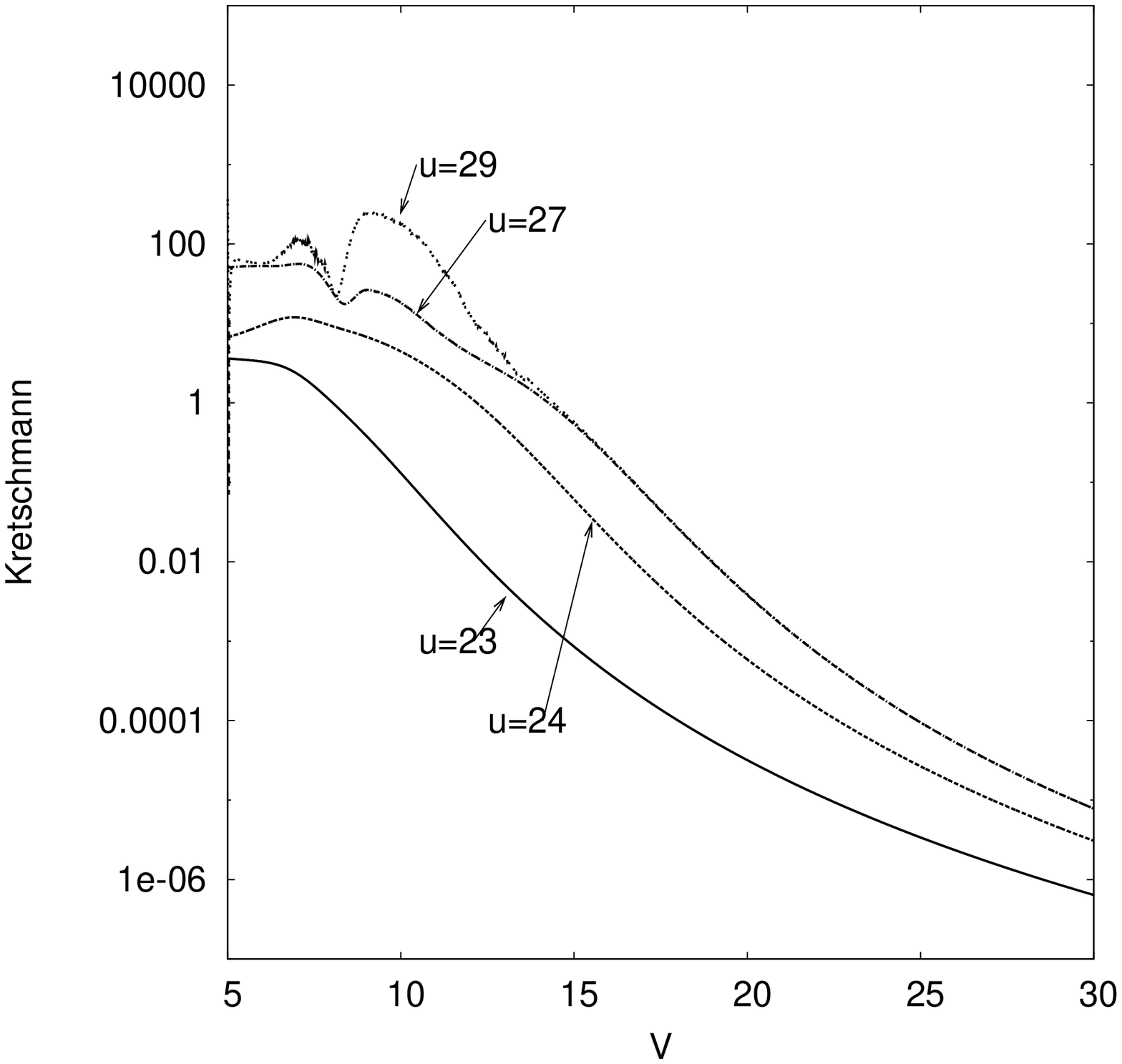}}
\caption{{\label{R10}  Kretschmann scalar along lines of constant
$u$ for the cases: a) $A_\Psi=0.05045$; b) $A_\Psi=0.0513$; c)
$A_\Psi=0.070$. In all cases $A_\Phi=0$, ($v_{\Psi 0}=5$, $v_{\Psi
1}=9$) and in all cases the BH is destroyed.}}
\end{figure}

\begin{figure}
\includegraphics[width=0.48\textwidth]{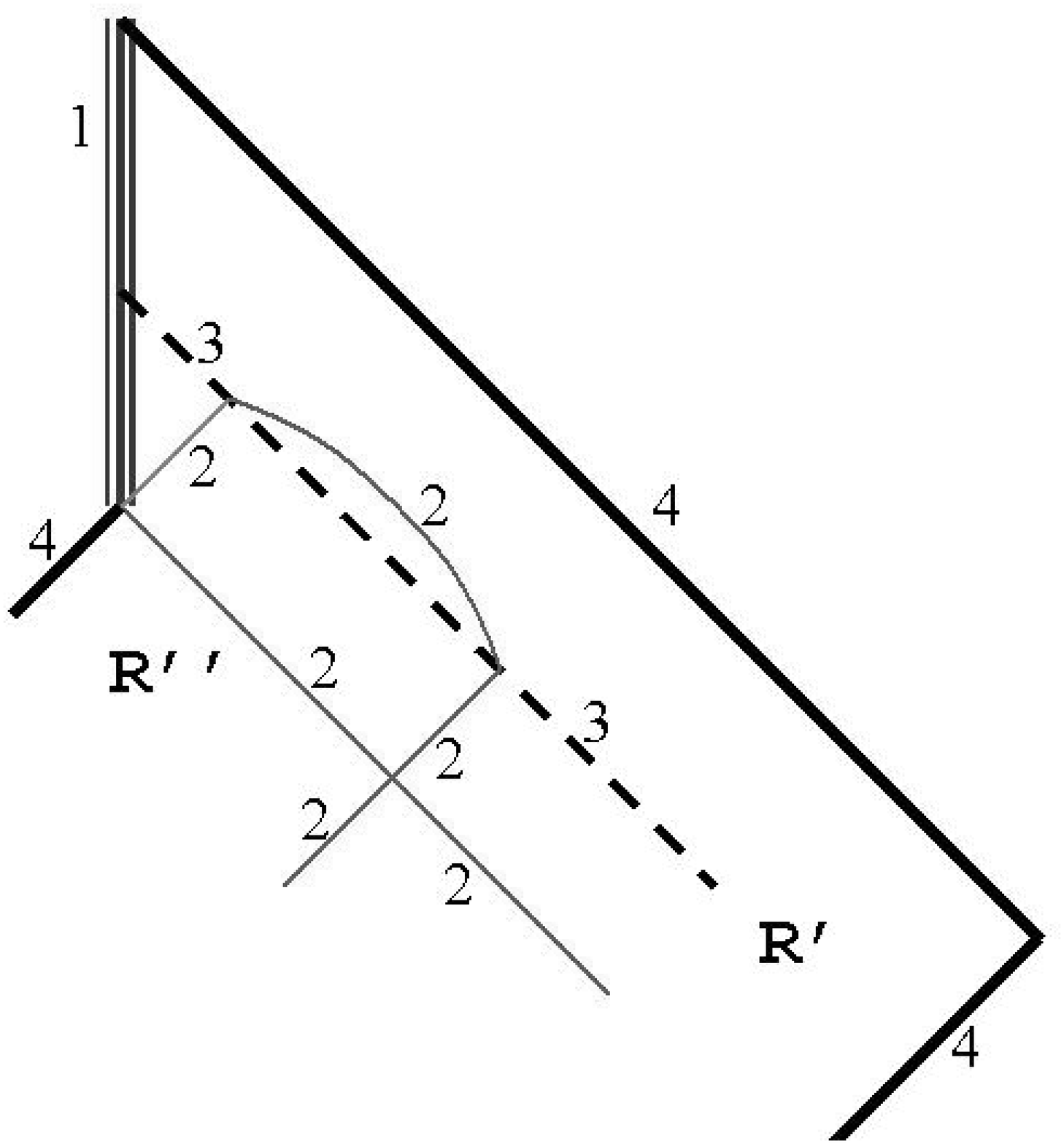}
\caption{{\label{R11} The Penrose diagram for the case when the BH
is destroyed after being irradiated by a pulse of the exotic scalar
field. Here: 1 -- strong singularity $r=0$, 2 -- horizons, 3 --
narrow pulse of the exotic scalar field, 4 -- light infinity, $R'$
-- our universe, $R''$ -- another universe.}}
\end{figure}

For all cases (\ref{ex5-a}-\ref{ex5-c}) the mass-function does not
demonstrate a mass inflation effect, see Fig.~\ref{R9}.
We note that for the cases (\ref{ex5-a}) and (\ref{ex5-b}) the
mass function for big $v$ and big $u$ does becomes greater than
${m=0.95}$. However, this is related to the scattering of the exotic scalar
field outside to bigger $r$. Of cause this is possible in the
case of dynamical BHs. In this region (big $v$ and big $u$) we
are at big $r$, definitely outside the (dynamical) BH which is at
smaller $v$.

For the cases (\ref{ex5-a}-\ref{ex5-b}) the Kretschmann scalar
does not increase with $v$, but rather oscillates (see
Figs.~\ref{R10}). These oscillations are the consequences of the
oscillations of the $T_{vv}$-in flux (and also $T_{uu}$-out flux),
which we described for the pulses with the smaller power, see
Fig.~\ref{R5a}-\ref{R5b}.

The fact that the Kretschmann scalar does not increase
exponentially with $v$, together with the fact that the mass
function does not increase exponentially, indicates the absence of
space-time singularities (excluding the timelike $r=0$
singularity of the Reissner-Nordstr\"om BH beyond the top-left
corner of the computational domain).

Finally, in Fig.~\ref{R8c} is shown the horizon structure for the
case~(\ref{ex5-c}).

In this case, the exotic pulse is so powerful that the inner and
outer apparent horizons meet quite fast and after this, there are not any
horizons or borders left at all (to the right of $v\sim 7.5$). The
outgoing signal $u=const$ can freely propagate away through the
$R$-region. Also, the mass function becomes essentially smaller
then $q$ (see Fig.~\ref{R9c}), while the Kretschmann scalar
becomes smaller and smaller with bigger $v$ (see
Fig.~\ref{R10}(c)). There are not any singularities present (
except for the timelike $r=0$ singularity).

Fig.~\ref{R11} represents the Penrose diagram that is confirmed by
our numerical simulations for the cases when the BH is destroyed
by the radiation.

At the end of this section we note the following. When the
Reissner-Nordstr\"om BH is irradiated by a pulse of the exotic
scalar radiation, the OAH becomes smaller (or disappears completely)
and part of the outgoing radiation from the $T_-$ region can go to
the outer $R'$-region in our Universe. This radiation may come into the $T_-$
region from the $R''$-region that belongs to another Universe, which is the counterpart of the outer
$R'$-region of Fig.~\ref{R11} in our Universe (from the left hand side of
Fig.~\ref{R11} outside the computational domain). This means
that it is possible for some radiation from the other Universe to come to our
$R'$-region. The propagation of the radiation in the opposite
direction, from our $R'$-region to the $R''$-region in the other
Universe, is still impossible. We call such an object a
semitraversable wormhole~\cite{AstRep2009}.

\section{The case of the irradiation by both normal and exotic pulses} \label{sec9}
Finally we consider the case:
\begin{equation}\begin{split}
\label{case_fig_8}  &A_\Phi=0.4,\quad (v_{\Phi 0}=5,\quad v_{\Phi 1}=7)  \\
                    &A_\Psi=0.2,\quad (v_{\Psi 0}=10,\quad v_{\Psi 1}=12).
\end{split}
\end{equation}

\begin{figure}
\includegraphics[width=0.48\textwidth]{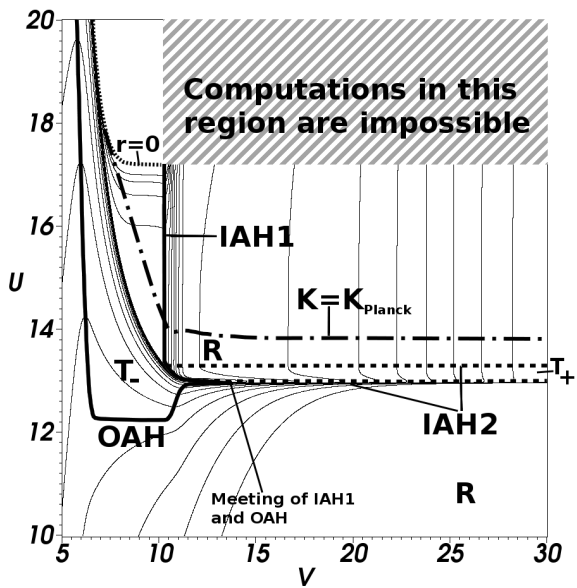}
\caption{Lines of constant $r$ (thin lines, decreasing from bottom to top), positions of $r=0$,
$K=K_{planck}$, apparent horizons and $R$- and $T$-regions for case: $A_\Phi=0.4$, ($v_{\Phi 0}=5$,
$v_{\Phi 1}=7$);
$A_\Psi=0.2$, ($v_{\Psi 0}=10$, $v_{\Psi 1}=12$) in which the BH is destroyed. \label{R12}}
\end{figure}

\begin{figure}
\includegraphics[width=0.48\textwidth]{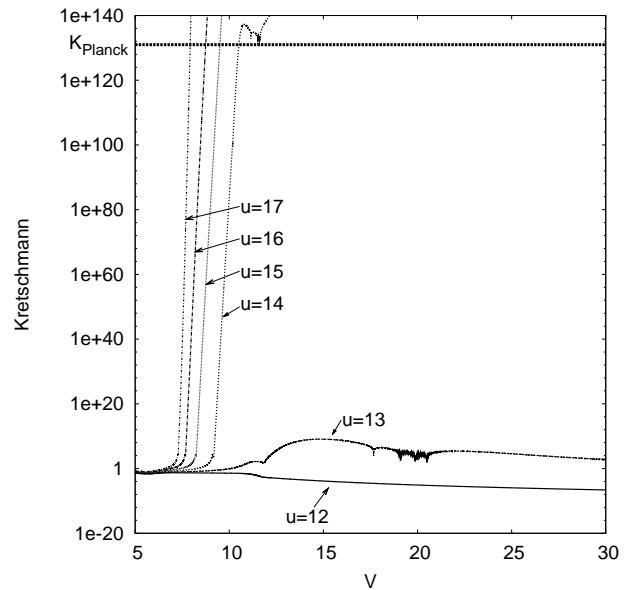}
\caption{Kretschmann
scalar versus $v$ along lines of constant $u$ for case: $A_\Phi=0.4$, ($v_{\Phi
0}=5$, $v_{\Phi 1}=7$);$A_\Psi=0.2$
($v_{\Psi 0}=10$, $v_{\Psi 1}=12$). Also, the location of the Kretschmann singularity $K_{planck}$ is plotted.\label{R13}}
\end{figure}

In this case, the Reissner-Nordstr\"om BH is irradiated at the
beginning by a strong pulse of the normal scalar field which causes a strong
spacelike $r=0$ singularity to arise in the $T_{-}$-region inside
the BH (similar to the cases investigated in~\cite{Hansen1}). After that, at higher $v$, the BH is then irradiated by a
strong pulse of the exotic scalar field.

For this case, one can see strong nonlinear processes.
Fig.~\ref{R12} shows the general structure of the apparent horizons and the
position of the (spacelike) singularity $r=0$. There is also a line showing
the position of the Kretschmann scalar equal to its Planckian
value of $K=(G\hbar /c^3)^{-2}\approx 10^{131}cm^{-4}$. Beyond
this border the classical General Relativity is not applicable
and we consider this region as a singularity from the classical
point of view. Our calculations are thus not reliable for this region
and we will not consider it (upper-right part of
Fig.~\ref{R12}, above the line $K=K_{Planck}$). We would like to do
the following remark about the $K$-scalar. This scalar has
dimension $cm^{-4}$ in the CGS -units system. In the paper
\cite{Hansen1} and in this paper we represent all figures in
dimensionless units. For the figures, we draw the $K$-scalar (and
the value of the Planckian $K$-scalar) for the case when the
linear scale (the size of the BH in this paper) is equal to unity.

The Kretschmann scalar is represented in Fig.~\ref{R13}. For large
$u\ge 14$, the Kretschmann scalar grows catastrophically fast and
reach the singularity $K=K_{Planck}$. This is the consequence of
the influence of the first pulse, with the positive energy density.
This is the same process which we analyzed in~\cite{Hansen1},
see e.g. Fig.~13 in~\cite{Hansen1}.
Along $u=12$, the Kretchmann scalar decreases with increasing $v$, this
is of course as expected as this line is always outside of any horizon (fig.~\ref{R12}).
The behavior along the line $u=13$ is somewhat in between, for small $v$ ($v\lesssim 14$),
the Kretschmann scalar shows a slight tendency to increase. However, for larger $v$, the BH
is destroyed and the Kretchmann scalar shows a decreasing behaviour. Also cf. with the discussion
of the mass function below.

\begin{figure}
\includegraphics[width=0.48\textwidth]{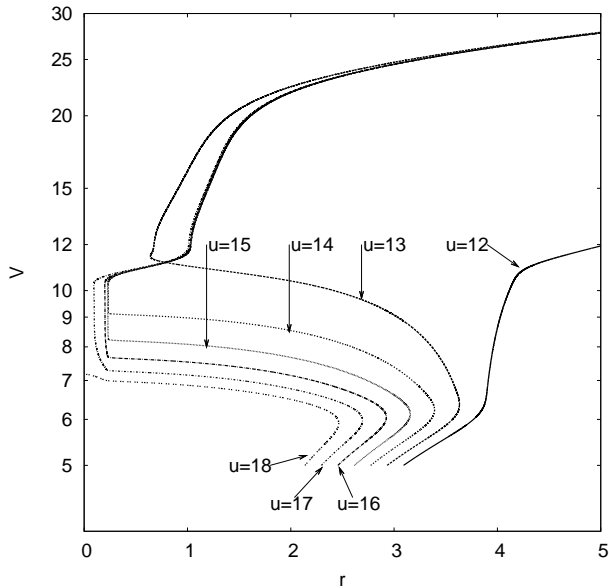}
\caption{$r$ versus $v$
along constant lines of constant $u$ for case: $A_\Phi=0.4$, ($v_{\Phi 0}=5$,
$v_{\Phi 1}=7$);
$A_\Psi=0.2$, ($v_{\Psi 0}=10$, $v_{\Psi 1}=12$). All lines for $u\ge 17$ can in principle escape
to infinity, except that most of them hit the Kretschmann singularity (see text for details). \label{R14}}
\end{figure}

In Fig.~\ref{R14} (showing $r$ versus $v$), the test photons along
$u=18$ reaches $r=0$. Photons in the range $13\lesssim u \lesssim
17$ come close to $r=0$ and after that, they propagate to larger
$r$ (after the meeting with the pulse of the exotic scalar field at
$10 \lesssim v \lesssim 12$). However, it is reminded that photons
along $u\gtrsim 14$ meet with $K=K_{Planck}$ and thus effectively reach
the singularity before they can escape to higher $r$. The photons
along $u=12$ constantly go to bigger $r$ because they are always
outside of any horizon.

\begin{figure}
\includegraphics[width=0.48\textwidth]{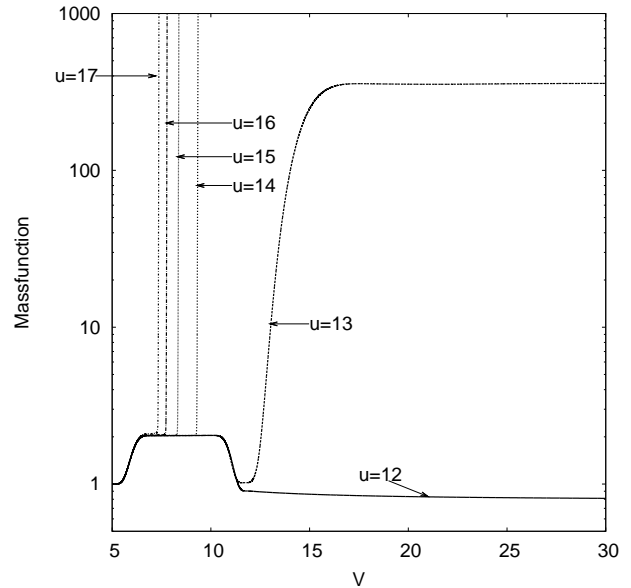}
\caption{Mass function along lines of constant $u$ for case: $A_\Phi=0.4$,
($v_{\Phi 0}=5$, $v_{\Phi 1}=7$);
$A_\Psi=0.2$, ($v_{\Psi 0}=10$, $v_{\Psi 1}=12$). Effects of both the normal and the exotic
scalar pulse are clearly seen. \label{R15}}
\end{figure}

In Fig.~\ref{R15}, it is seen that the mass function goes up for
all lines of $u=constant$ between $5\le v \le 7$, this is merely the trivial effect of mass being pumped into the BH by the positive energy pulse.
In addition to this, the mass function for $u \ge 14$ continues to increase even after the passage of the positive energy pulse, due to the processes
described in section~\ref{sec7} until they reach the $K=K_{Planck}$ singularity.

As the negative energy pulse goes into the BH, the mass function along lines $u \le 13$ decreases
due to the negative energy contents of the second pulse. This second pulse, causes the mass function
to become less than $q$ in the region near the outer horizon and it leads to the meeting of the
inner and outer apparent horizons and the disappearance of the BH around $v\approx 14$ (see figure ~\ref{R12}).
Along $u=12$, the mass function continues to decrease because of the scattering processes of the exotic scalar field.
Along $u=13$, however, we see that the mass function increases drastically near ${v\approx 12}$.
This growth is probably caused by the mass inflation effect, since in this region we are close to the border
of the BH, but still inside it. The apparent horizons do not meet and the BH is not destroyed until around $v\approx 14$.
In the same region, one can also see a modest growth of the Kretschmann scalar in Fig.~\ref{R13}. However,
as the BH is destroyed around $v\approx 14$, we see that the mass function halts its growth and the Kretschmann
scalar starts to decrease.

\section{Conclusions}
\label{concl}
The processes arising when a Reissner-Nordstr\"om BH is
irradiated by a pulse of an exotic scalar field with a negative energy
density have been analyzed. We have performed the corresponding
numerical computations using a numerical code specially designed for the purpose. It was
demonstrated that these processes are quite different from the
processes arising in the case of the irradiation of a
Reissner-Nordstr\"om BH by a pulse of a normal scalar field.

The evolution of the mass function and the Kretschmann scalar
demonstrate that in the case of the exotic scalar field, the
evolution does not lead to the origin of a strong spacelike
singularity $r=0$ in the $T$-region as was the seen in the
case of irradiation by the normal scalar field~\cite{Hansen1}.

The numerical calculations demonstrate the manifestation of the
antifocusing effects in the gravity field of an exotic scalar
field with a negative energy density.

When the power of the exotic pulse with negative energy density is
great enough, the mass function becomes less than the charge $q$ near the outer apparent horizon.
As a result the BH disappears. This process was analyzed in detail.

\section*{Acknowledgements}
\label{Acknowledgements}
We thank the unknown referees of our paper for very
constructive criticism and helpful advice. This work was supported
in part by the JSPS Postdoctoral Fellowship For Foreign
Researchers, the Grant-in-Aid for Scientific Research Fund of the
JSPS ${(19-07795)}$, Russian Foundation for Basic Research
(project codes: $07-02-01128-a$, $08-02-00090-a$,
$08-02-00159-a$), scientific schools: $NSh-626.2008.2$,
$Sh-2469.2008.2$ and by the program {\it Origin and Evolution of
Stars and Galaxies 2009} of the Russian Academy of Sciences.

\appendix
\section{Convergence of the code}
\label{sec:app1}
In this appendix, we demonstrate that our numerical code is
converging (to a physical solution) when including the effects of
nontrivial $\Phi$ and $\Psi$ fields. The results presented here
are very similar to those presented in \cite{Dorr2} as could be
expected, since the codes in that paper and this are essentially
identical. For this reason, for further details of the code, we
refer to \cite{Dorr2}.

The initial conditions for the tests in this Appendix are similar
to those described in section \ref{sec9}, i.e. the initial
conditions are those of a Reissner-Nordstrom BH, perturbed by
first an in-falling $\Phi$ field (from $v=[5;7]$ with amplitude:
$A_{\Phi}=0.4$) and subsequently by an in-falling $\Psi$ field
(from $v=[10;12]$ with amplitude: $A_{\Psi}=0.2$). Our
computational domain for the convergence tests are, as for all
simulations in this paper, in the range $v=[5;30]$ and $u=[0;30]$.
This setup is the most complicated one that we have done in this
paper (in that it incorporates the dynamic effects of both $\Phi$
and $\Psi$ fields) and leaves no trivial terms left in the evolution
equations, thus it is a good test for the convergence of our code.

Here we demonstrate the convergence of the code by comparing a
series of simulations with varying numerical resolution. For the
tests in this Appendix, we do a total of 6 simulations, with each
simulation changing the AMR base resolution and refinement criteria
in such a way as to mimic doubling the numerical resolution from
one simulation to the next (for further details and more thorough
discussions, see \cite{Pretorius04, Dorr2}).

To limit the number of plots in this Appendix, we concentrate on
displaying convergence results along the line $u=17$ (however, it
should be noted that a large number of convergence tests has been
carried out and that the results presented in this Appendix are
representative of the convergence properties of the code in other
parts of the computational domain and for other configurations of
the initial data). As can be seen in figure \ref{R12}, this line is
very close to the $r=0$ singularity (where we can expect strong
dynamics) as well as inside the Kretschmann singularity and thus convergence along this line is a good
indication of the convergence behavior throughout the computational domain.

Along this line, we calculate the relative convergence between two simulations (one with a numerical resolution twice that of the other) relative to
a simulation with very high resolution:
\begin{equation}
  \label{eq:xidef}
  \xi (x_N^i) \equiv \frac{|x_N^i - x_{2N}^i|}{ |x_{HighRes}^i |}
\end{equation}
where $x_N^i$ denotes the dynamic variable $x$ at the $i$-th grid
point of simulation with resolution $N$ and where $x_{HighRes}^i$
denotes the dynamic variable of the same $i$'th point for a
simulation with the highest numerical resolution done by us.
Obviously, this expression only makes sense for those $i$ points
that coincide in all simulations.

\begin{figure*}
\subfigure[ Relative convergence of $r$.
\label{fig:A1a}]{\includegraphics[width=0.45\textwidth, height=0.38\textwidth]{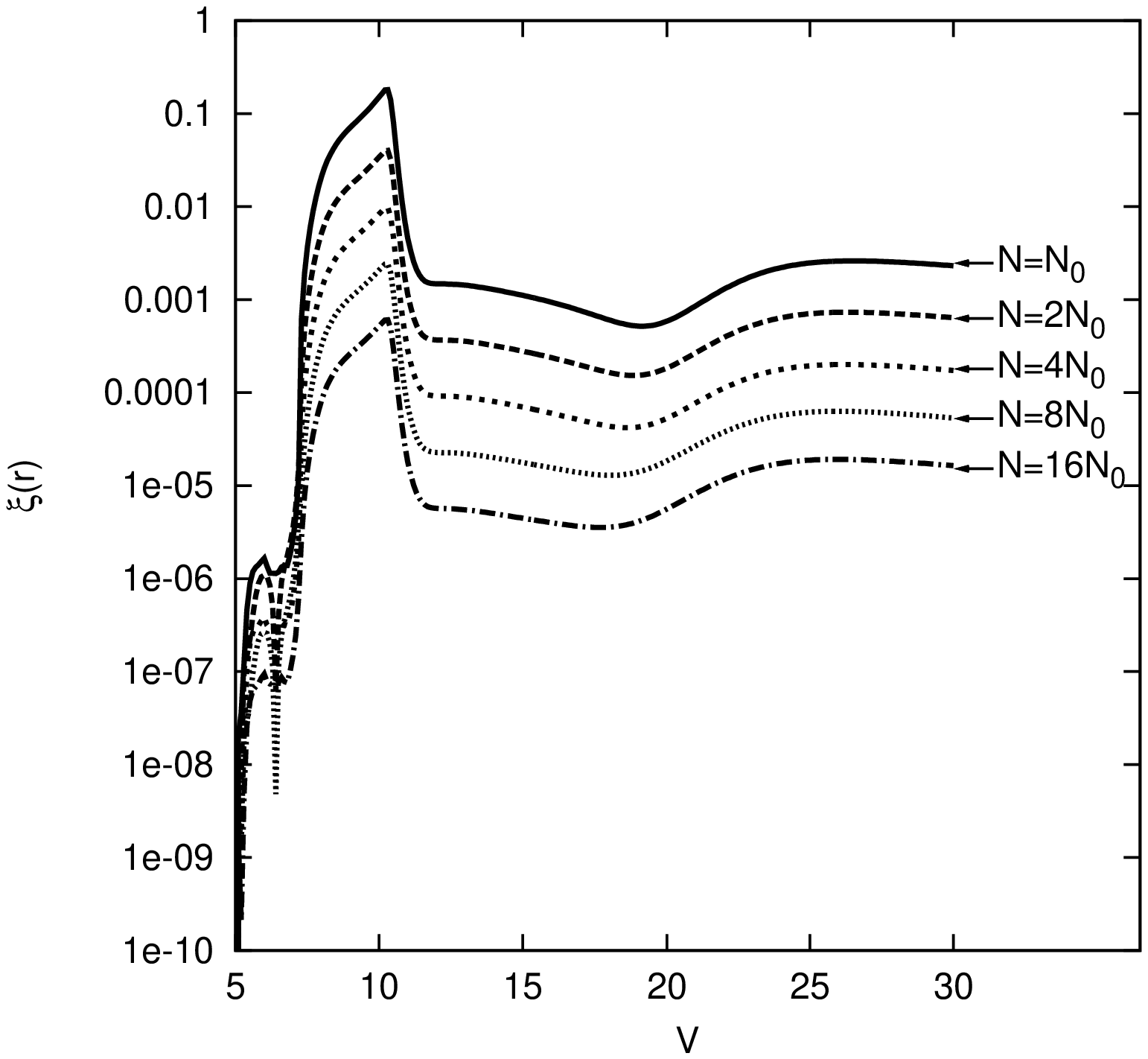}}
\subfigure[ Relative convergence of $\sigma$.
\label{fig:A1b}]{\includegraphics[width=0.45\textwidth, height=0.38\textwidth]{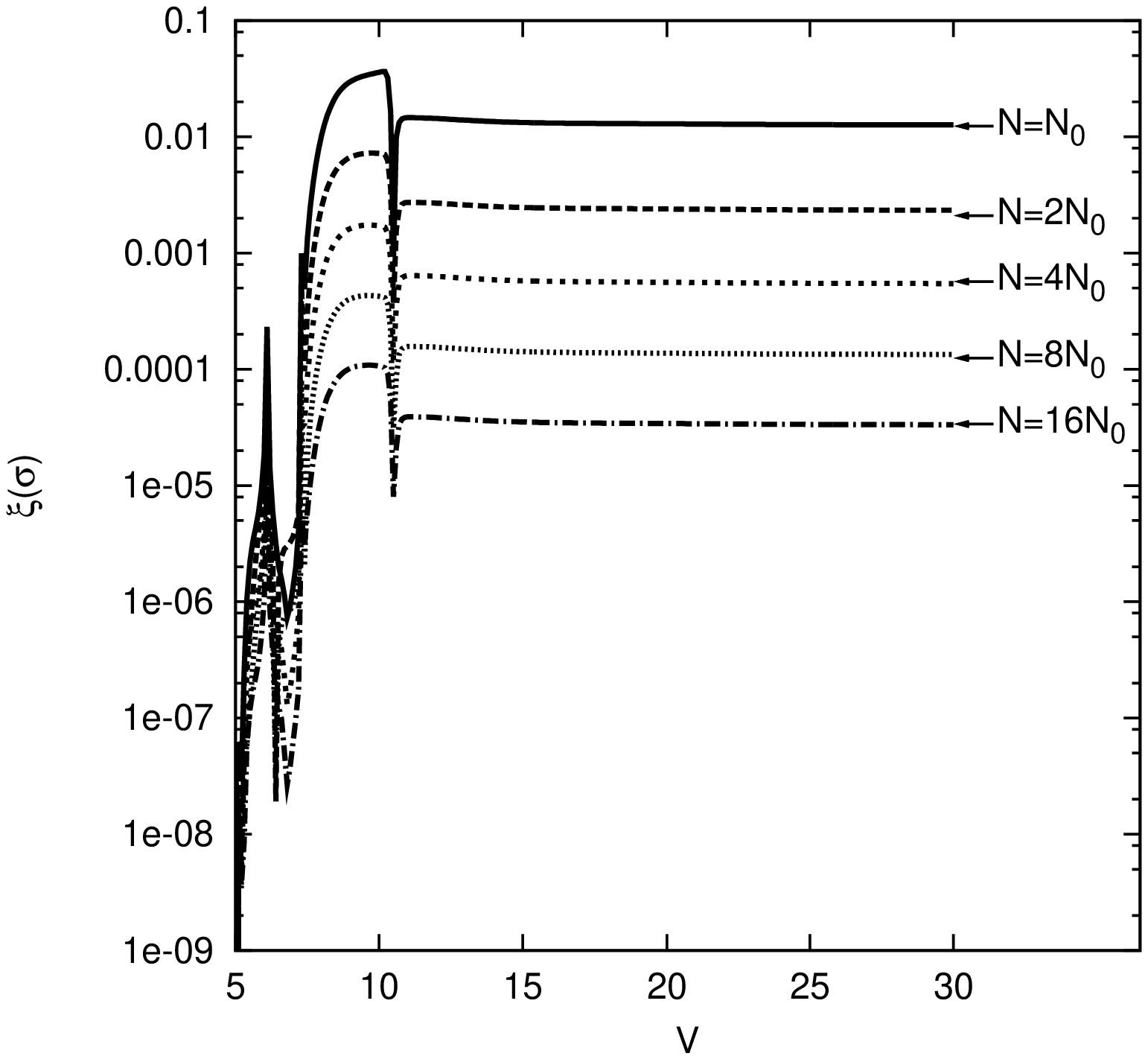}}
\subfigure[ Relative convergence of $\Phi$.
\label{fig:A1c}]{\includegraphics[width=0.45\textwidth, height=0.38\textwidth]{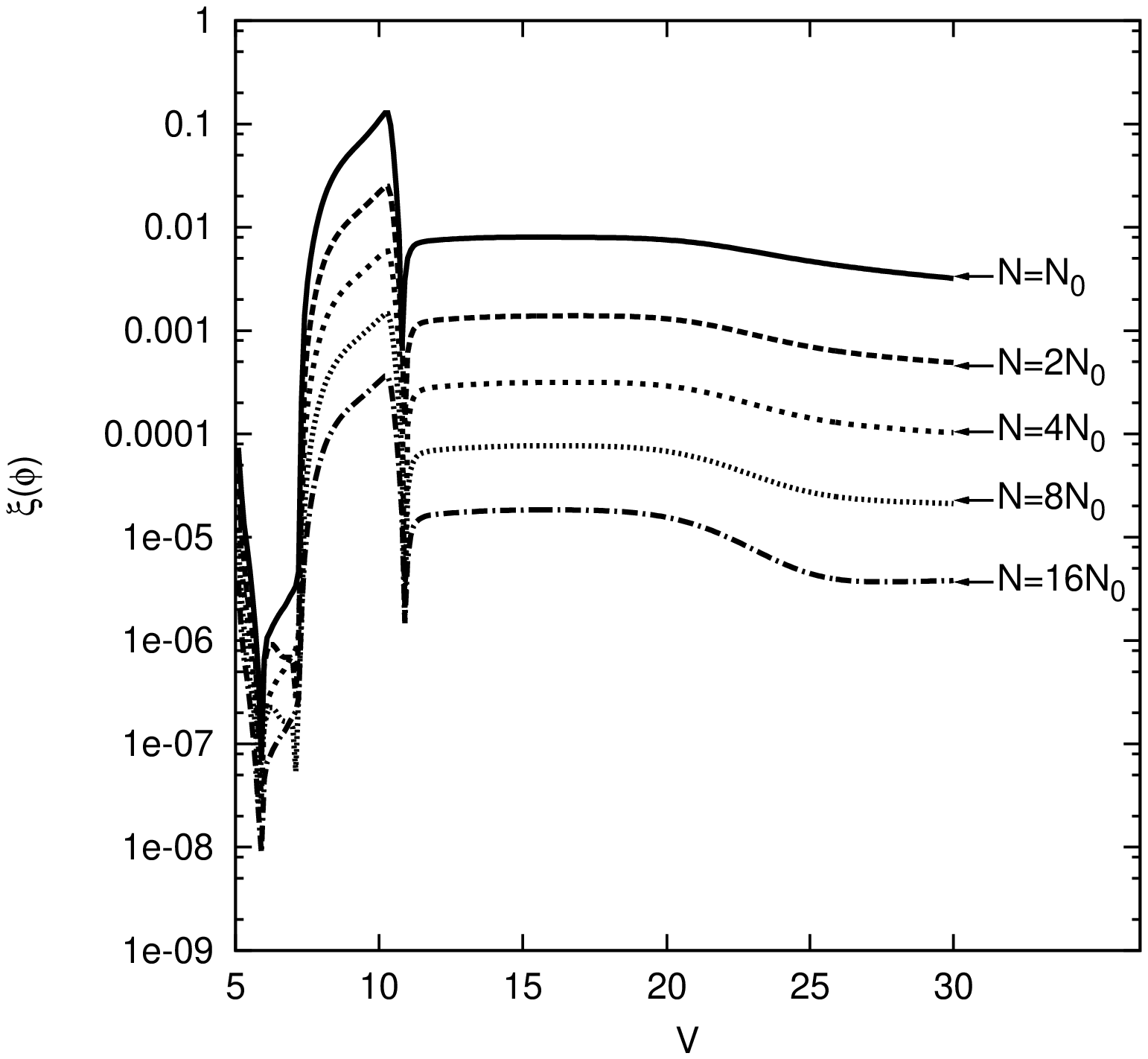}}
\subfigure[ Relative convergence of $\Psi$.
\label{fig:A1d}]{\includegraphics[width=0.45\textwidth, height=0.38\textwidth]{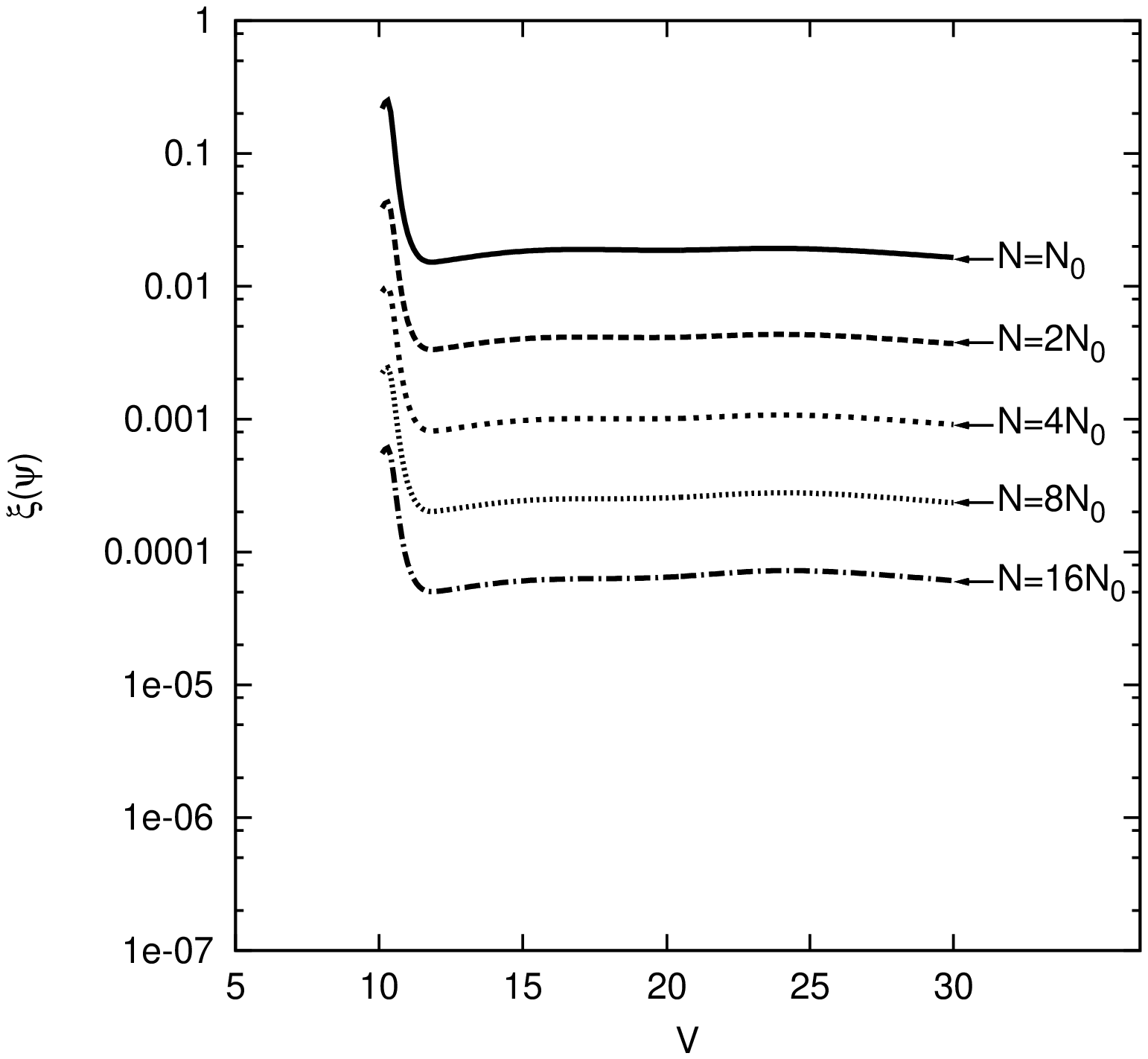}}
\subfigure[ Relative convergence of $C_{uu}$.
\label{fig:A1e}]{\includegraphics[width=0.45\textwidth, height=0.38\textwidth]{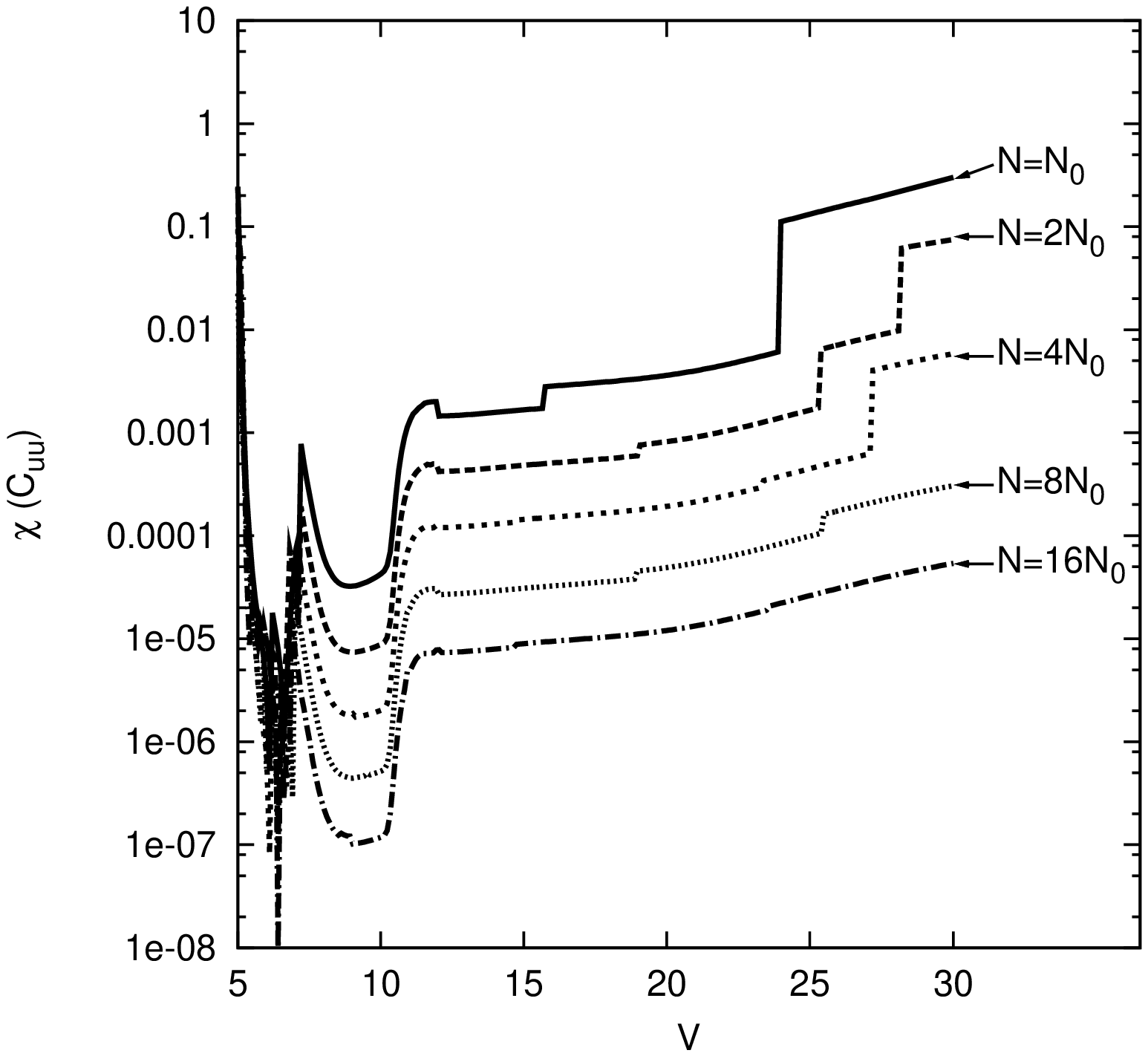}}
\subfigure[ Relative convergence of $C_{vv}$.
\label{fig:A1f}]{\includegraphics[width=0.45\textwidth, height=0.38\textwidth]{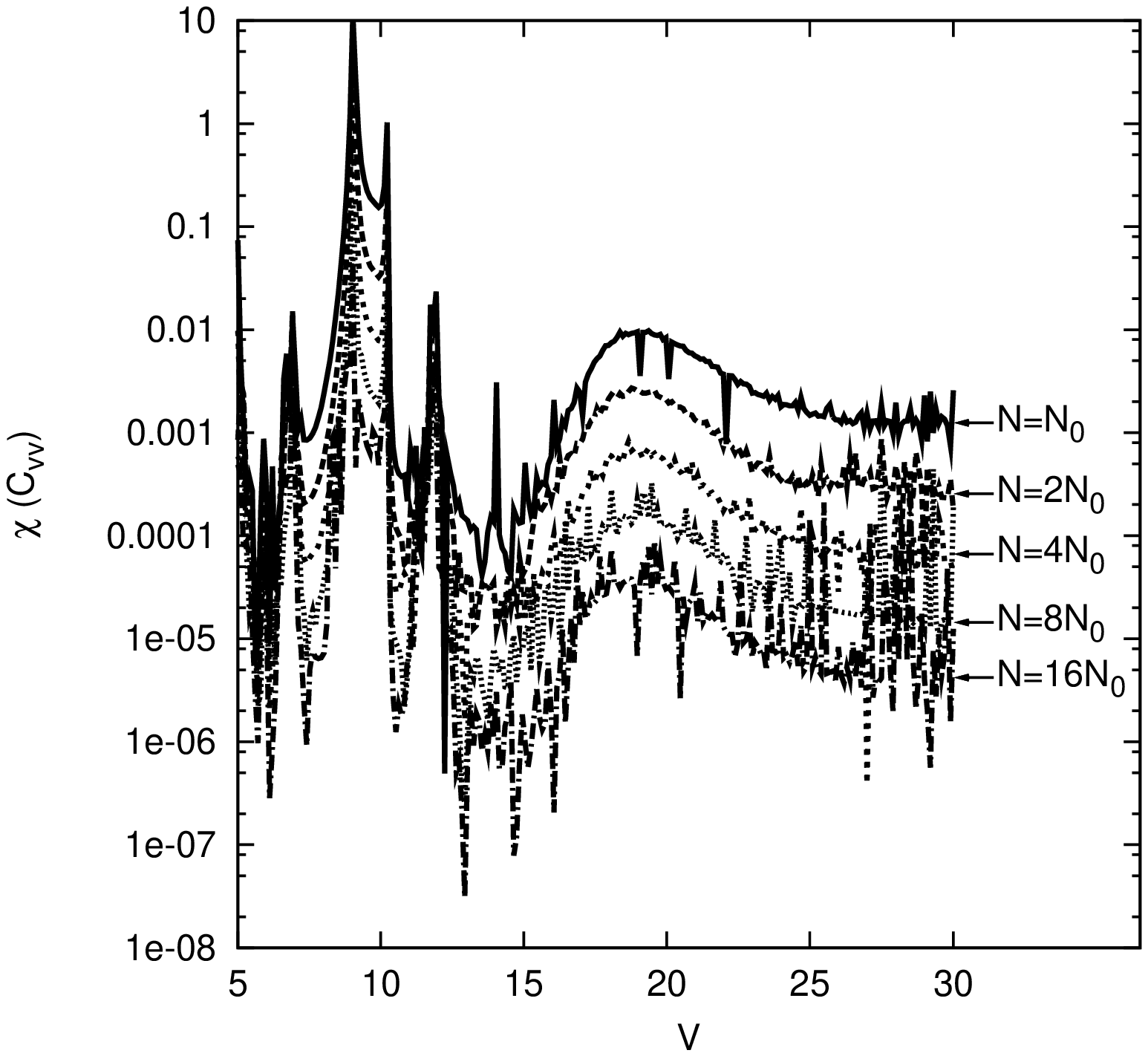}}
\caption{{\label{fig:A1} Relative convergence of the dynamic variables and constraint
equations along line of $u=17$, see text for details.}}
\end{figure*}
Figures \ref{fig:A1a}-\ref{fig:A1d} show the relative convergence,
$\xi$, for the dynamic variables $r,\sigma,\Phi$ and $\Psi$
respectively. The lines in the figures are marked by their numerical resolution measured in terms of the most coarse resolution $N_0$, the high resolution simulation
used to calculate expression \eqref{eq:xidef}, has a numerical resolution of $32$ times the base resolution, i.e. $N=32N_0$.

From these figures, it is clearly seen that the four
dynamic variables are converging for simulations of increasing
resolution. Furthermore, since we plot the \textit{relative}
convergence of the dynamic variables, we see that the relative
change between the two highest resolution simulations show that
the variables change $0.1\%$ or less, which must be considered a
quite acceptable convergence. We note that a closer analysis of the data in
figures \ref{fig:A1a}-\ref{fig:A1d} has revealed that the dynamic
variables are indeed converging with second order accuracy as was
expected based on analyses and tests performed in
\cite{Dorr2,Hansen1}.

However, of course it is not enough to demonstrate that the
simulations are converging, they must also converge to a physical
solution, i.e. the residuals of the constraint equations (eqs.
\eqref{eq:7} and \eqref{eq:8}) must converge to zero. To
demonstrate this, we calculate the relative convergence of the
constraint equation residuals, (relative to the Einstein-tensor,
eqs. \eqref{eq:Guu} and \eqref{eq:Gvv} respectively) in a similar way to eq. \eqref{eq:xidef} :
\begin{equation}
  \label{eq:chidef}
  \chi (C_N^i) \equiv \frac{|C_N^i|}{ |G_{HighRes}^i |}
\end{equation}
where $C_N^i$ denotes the residual of the constraint equation,
($C_{uu}$ or $C_{vv}$), at the $i$'th point for simulation with
resolution $N$ and where $G_{HighRes}^i$ denotes the corresponding
Einstein-tensor component ($G_{uu}$ or $G_{vv}$ respectively) at
the same point.

The relative convergence of the residuals of the constraint
equations are demonstrated in figures \ref{fig:A1e} and
\ref{fig:A1f} where it is seen that they converge towards zero for
higher resolution simulations. This indicates that not only are
the numerical solutions converging for simulations of higher
resolution, but that they are indeed converging towards a physical
solution. The convergence plots for the constraint equations
exhibit significant more numerical noise than the convergence
plots for the basic dynamic variables. However, it is reminded
that the constraint equations are derived quantities of the basic
dynamic variables, therefore the constraint equations will be less
accurate and include more numerical errors such as truncation
errors from calculating the derivatives of the basic dynamic
variables and errors associated with calculating derivatives in
the AMR mesh. Nevertheless, figs. \ref{fig:A1e} and \ref{fig:A1f}
indicate that the solutions are indeed converging to the physical
solutions for high resolutions.

Finally it should be noted that the convergence results presented
in this appendix are not the only convergence tests that we have
performed, they merely represent typical results of the
convergence behavior of the code. For all results presented in
this paper, we have performed a large number of simulations with
varying resolutions to ensure that the results had converged to
their physical solution.

\bibliography{paper}

\end{document}